\documentclass[aps,prstab,preprint,groupedaddress]{revtex4}
\usepackage{graphicx}

\begin{document}

\title{Crowd-Anticrowd Theory of Collective Dynamics in Competitive,
Multi-Agent Populations and Networks}


\author{Neil F. Johnson$^*$}
\affiliation{Physics Department, Oxford University, 
Oxford, OX1 3PU, UK}
\author{Pak Ming Hui}
\affiliation{Department of Physics, The Chinese University of Hong Kong, Shatin, Hong Kong,
China}

\date{\today}

\begin{abstract}
We discuss a crowd-based theory for describing the collective behavior in a generic multi-agent
population which is competing for a limited resource. These systems -- whose binary versions we refer
to as B-A-R (Binary Agent Resource) collectives -- have a dynamical evolution which is determined by
the aggregate action of the heterogeneous, adaptive agent population. Accounting for the strong
correlations between agents' strategies, yields an accurate description of the system's dynamics in
terms of a `Crowd-Anticrowd' theory. This theory can incorporate the  effects of an underlying
network within the population. Most importantly, its applicability is {\em not} just limited to the El
Farol Problem and the Minority Game. Indeed, the Crowd-Anticrowd theory  offers a
powerful approach to tackling the dynamical behavior of a wide class of agent-based Complex Systems,
across a range of disciplines. With this in mind, the present working paper is written for a general
multi-disciplinary audience within the Complex Systems community.
\vskip0.5in
\noindent {\bf Working paper for the Workshop on Collectives and the Design of Complex Systems,
Stanford University, August 2003.}

\noindent {$^*$ Research performed in collaboration with former graduate students Michael Hart and
Paul Jefferies, and present graduate students Sehyo Charley Choe, Sean Gourley and David Smith.}

\end{abstract}

\maketitle

\section{Introduction}
Complex Systems -- together with their dynamical behavior known as Complexity -- are thought to
pervade much of the natural, informational, sociological, and economic world
\cite{casti1,casti2,sole,bossmaker,holland}. A unique, all-encompassing definition of a
Complex System is lacking - worse still, such a definition would probably end up being 
too vague. Instead, such Complex Systems are better thought of in terms of a list of
common features which distinguish them from `simple' systems, and from systems which are
just `complicated' as opposed to being complex. Although a unique list of Complex
System properties does not exist, most people would agree that the following would typically
appear: feedback and adaptation at the macroscopic and/or microscopic level, many (but not too
many) interacting parts, non-stationarity, evolution, coupling with the environment, and
observed dynamics which depend upon the particular realization of the system. In addition,
Complex Systems have the ability to produce large macroscopic changes which appear spontaneously but
have long-lasting consequences. Such large changes might also be referred to as `innovations',
`breakthroughs', `gateway events' or `punctuated equilibria' depending on the context
\cite{gellman}. Alternatively, the particular trajectory taken by a Complex System can
be thought of as exhibiting `frozen accidents' \cite{gellman}. Understanding the
functionality of Complex Systems is of paramount importance, from both practical and theoretical
viewpoints. Such functionality is currently being addressed through the study of `Collectives'
\cite{davidnasa}.

The down-side of labelling such a wide range of systems as belonging to
the same `Complex' family, is that instead of saying something about everything, one may end
up saying nothing very much about anything. Indeed, one may end up with little more than the
vague notion that `ant colonies are like vehicular traffic, which is like
financial markets, which are like fungal colonies etc.'.
On the other hand, it would be a mistake to focus too narrowly on a specific example of a Complex
System since the lessons learned may not be transferable - worse still, they may be misleading or
plain wrong in the context of other Complex Systems. Such is the daunting task facing researchers in
Complex Systems: based on studies of a few very specific Complex System models, how
can one extract general theoretical principles which have wide applicability across a
range of disciplines? This explains why, to date, there are very few truly universal
theoretical principles or `laws' to describe Complex Systems.

As pointed out by John Casti on p. 213 of Ref. \cite{casti1}, `.... a decent mathematical
formalism to describe and analyze the [so-called] El Farol Problem would go a long way toward
the creation of a viable theory of complex, adaptive systems'. The rationale behind this
statement is that the El Farol Problem, which was originally proposed by Brian
Arthur\cite{farol} to demonstrate the essence of Complexity in financial markets,
incorporates the key features of a Complex System in an everyday setting. Very
briefly, the El Farol Problem concerns the collective decision-making of a group
of potential bar-goers, who repeatedly try to predict whether they should attend a
potentially overcrowded bar on a given night each week. They have no
information about the others predictions. Indeed the only information available to each agent is
global, comprising a string of outcomes (`overcrowded' or `undercrowded') for a limited number of
previous occasions. Hence they end up having to predict the predictions of others. No `typical' agent
exists, since all such typical agents would then make the same decision, hence rendering their common
prediction scheme useless. This simple yet intriguing problem has inspired a huge amount of interest
in the physics community over the past few years. Reference
\cite{uselfarol}, which was the first work on the full El Farol Problem in the physics community,
identified a minimum in the volatility of attendance at the bar with increasing adaptivity of the
agents. With the exception of Ref. \cite{uselfarol}, the physics literature has instead focussed on a
simplified binary form of the El Farol Problem as introduced by Challet and Zhang
\cite{origMG}. This so-called Minority Game (MG) is discussed in detail in Refs.
\cite{RSS,ChalletContTime,ChalletPhasTran,ChalletSpin,ChalletStyle,
marsili2,marsiliThermo,memory,comment,SavitLet,SavitPhysA,
SavitPhysAb,CAT1,CAT2,alloy,generalTherm,CAA,MarketModel,
THMG1,THMG2,paul,dublin,prl,heimel,heimel2,cavagna,SherThermLett, SherMarkMod}) and Ref.
\cite{webpage}. The Minority Game concerns a population of $N$\ heterogeneous agents with
limited capabilities and information, who repeatedly compete to be in the minority group. The
agents (e.g. people, cells, data-packets) are adaptive, but only have access to global
information about the game's progress. In both the El Farol Problem and the Minority Game, the
time-averaged fluctuations in the system's global output are of particular importance --
for example, the time-averaged fluctuations in attendance can be used to assess the wastage
of the underlying global resource (i.e. bar seating).

Despite its appeal, the El Farol Problem (and the Minority Game in particular) is
somewhat limited in its applicability to general Complex Systems, and hence arguably
falls short of representing a true generic paradigm for Complex Systems. First, the reward structure
is simultaneously too restrictive and specific. Agents in the Minority Game, for example, can
only win if they instantaneously choose the minority group at a particular timestep: yet this is
not what investors in a financial market, for example, would call `winning' \cite{book}. Second,
agents only have access to global information. In the natural world, most if not all biological and
social systems have at least some degree of underlying connectivity between the agents, allowing for
local exchange of information or physical goods \cite{networks}. In fact it is this interplay of
network structure, agent heterogeneity, and the resulting functionality of the overall system, which
is likely to be of prime interest across disciplines in the Complex Systems field. 

Three enormous
challenges therefore face any candidate Complex Systems theory: it must be able to explain the
specific numerical results observed for the El Farol Problem and its variants; it should also be able
to account for the presence of an arbitrary underlying network; yet it should be directly
applicable to a much wider class of `game' with a variety of reward structures. Only then
will one have a hope of claiming some form of generic theory of Complex Systems. And only then will
the design of multi-agent Collectives to address both forward and inverse
problems, become relatively straightforward. It might then be possible to achieve the
Holy Grail of predicting {\em a priori} how to engineer agents' reward structures, or which specific
game rules to invoke, or what  network communication scheme to introduce, in order to achieve some
global objective. For a detailed discussion of these issues, we refer to the paper of Kagan Tumer
and David Wolpert \cite{davidnasa} at this Workshop. 

In this paper, we attempt to take a step in this
more general direction, by building on the success of the Crowd-Anticrowd theory in describing
both the original El Farol Problem \cite{uselfarol}, and the Minority Game \cite{CAT1,CAT2} and
its variants. In particular,  we present a formal treatment of the collective
behavior of a generic  multi-agent population which is competing for a limited resource. The
applicability of the Crowd-Anticrowd analysis is {\em not} limited to MG-like games, even though we
focus on MG-like games in order to demonstrate the accuracy of the Crowd-Anticrowd approach in
explaining the numerical results. We also  show how
the Crowd-Anticrowd theory can be extended to incorporate the presence of networks.  The theory is
built around the crowding (i.e. correlations) in strategy-space, rather than the precise rules of
the game itself, and only makes fairly modest assumptions about a game's dynamical behavior.
To the extent that a given Complex System mimics a general competitive multi-agent
game, it is likely that the Crowd-Anticrowd approach will therefore be applicable. This would
be a welcome development, given the lack of general theoretical concepts in the field of
Complex Systems as a whole. The challenge then moves to understanding how best to embed the
present theory within the powerful COIN framework developed by Wolpert
and Tumer for Collectives \cite{davidnasa}.

We have tried to aim this paper at a
multi-disciplinary audience within the general Complex Systems community. The layout of the
paper is as follows. Section II briefly discusses the background to the Crowd-Anticrowd framework.
Section III provides a description of a wide class of Complex Systems which will be our focus: we
call these B-A-R (Binary Agent Resource) problems in recognition of the stimulus provided by Arthur's
El Farol Problem. The Minority Game is a special limiting case of such B-A-R problems. Section IV
develops the Crowd-Anticrowd formalism which describes a general B-A-R system's dynamics. Section V
considers the implementation of the Crowd-Anticrowd formalism, deriving analytic expressions for the
global fluctuations in the system in various limiting cases. Section VI applies these results to
both the basic Minority Game and several generalizations, in the absence of
a network. Section VII then considers the intriguing case of B-A-R systems subject to an underlying
network. In particular, it is shown analytically that such network structure may have important
benefits at very low values of network connectivity, in terms of reduced wastage of global resources.
The conclusion is given in Section VIII.

\section{Crowd-Anticrowd framework}
The fundamental idea behind our crowd-based approach describing the dynamics of a complex
multi-agent system, is to incorporate accurately the correlations in strategies followed by the
agents. This methodology is well-known in the field of many-body theory in physics, particularly in
condensed matter physics where composite `super-particles' are typically considered which
incorporate all the strong correlations in the system. A well-known example is an exciton gas: each
exciton contains one negatively-charged electron and one positively-charged hole. Because of
their opposite charges, these two particles move in opposite directions in a given electric field
and attract each other strongly. However since this exciton super-particle is neutral overall,
any two excitons will have a negligible interaction and hence the excitons move independently of each
other. More generally, a given exciton (or so-called `excitonic complex') may contain 
a `crowd' of $n_e\geq 1$ electrons, together with a `crowd' of $n_h\geq 1$ holes having opposite
behavior. In an analogous way, the
Crowd-Anticrowd theory in multi-agent games forms groups containing like-minded agents (`Crowd')
and opposite-minded agents (`Anticrowd'). This is done in such a way that the strong
strategy correlations are confined within each group, leaving weakly interacting Crowd-Anticrowd
groups which then behave in an uncorrelated way with respect to each other. The first
application\cite{uselfarol} of our crowd-based approach was to the full El Farol Problem. It yielded
good quantitative agreement with the numerical simulations. However the analysis was complicated by
the fact that the space of strategies in the El Farol Problem is difficult to quantify.  This
analysis becomes far easier if the system has an underlying binary structure, as we will see in
Sections III and IV. However we note that the Crowd-Anticrowd analysis is not in principle limited
to such binary systems.

As indicated above, the Crowd-Anticrowd analysis breaks the
$N$-agent population down into groups of agents, according to the correlations between these
agents' strategies. Each group $G$ contains a crowd of agents using
strategies which are positively-correlated, and a complementary anticrowd using strategies
which are strongly negatively-correlated to the crowd. Hence a given group
$G$ might contain $n_R[t]$ agents who are all using strategy $R$ and hence act
as a crowd (e.g. by attending the bar en masse in the El Farol Problem) together with $n_{\overline
R}[t]$ agents who are all using the opposite strategy $\overline R$ and hence act as an anticrowd
(e.g. by staying away from the bar en masse). Most importantly, the anticrowd $n_{\overline R}[t]$
will always take the {\em opposite} decisions to the crowd $n_{R}[t]$ {\em regardless} of 
the current circumstances in the game, since the strategies
$R$ and $\overline R$ imply the opposite action in all situations. Note that this collective
action may be entirely involuntary in the sense that  typical games will be competitive, and there may
be very limited (if any) communication between agents.  Since all the strong correlations have been
accounted for within each group, these individual groups
$\{G\}=G_1,G_2,\dots,G_n$ will then act in an uncorrelated way with respect to each other, and
hence can be treated as $n$ uncorrelated stochastic processes. The global dynamics of the system is
then given by the sum of the $n$ uncorrelated stochastic processes generated by the groups
$\{G\}=G_1,G_2,\dots,G_n$.

Regarding the special limiting case of the Minority Game (MG), we note that there have 
been various alternative theories
proposed to describe the
MG's dynamics \cite{ChalletContTime,ChalletPhasTran,ChalletSpin,ChalletStyle,
marsili2,memory,heimel,heimel2,SherThermLett, SherMarkMod}. Although elegant
and sophisticated, such theories have however not been able to reproduce the original
numerical results of Ref.
\cite{SavitLet} over the full range of parameter space. What is missing from
such theories is an accurate description of the correlations between agents' strategies: in
essence, these correlations produce a highly-correlated form of decision noise which cannot
easily be averaged over or added in. By contrast these strong correlations take
center-stage in the Crowd-Anticrowd theory, in a
similar way to particle-particle correlations taking center-stage in many-body physics.

\section{B-A-R (Binary Agent Resource) systems}
\subsection{Generic B-A-R system setup}
Figure 1 summarizes the generic form of the B-A-R (Binary Agent Resource) system under
consideration. At timestep
$t$, each agent (e.g. a bar customer, a commuter, or a market agent) decides whether to
enter a game where the choices are action $+1$ (e.g. attend the bar, take route A, or buy) and
action $-1$ (e.g. go home, take route B, or sell). We will denote the number of agents choosing 
 $-1$ as $n_{-1}[t]$, and the number choosing 
 $+1$ as $n_{+1}[t]$. For simplicity, we will assume here that all the
agents participate in the game at each timestep, although this is not a necessary restriction. We can
define an `excess demand' as
\begin{equation} D[t]=n_{+1}[t]-n_{-1}[t] \ \ . \label{vol}
\end{equation}
As suggested in Figure 1, the agents may have connections between them: these connections
may be temporary or permanent, and their functionality may be conditional on some other features in
the problem.  The
global information available to the agents is a common memory of the recent history, i.e. the most
recent $m$ global outcomes. For example for 
$m=2$, the possible forms are $\dots 00$,
$\dots 01$,
$\dots 10$ or
$\dots 11$ which we denote simply as $00$, $01$, $10$ or $11$. Hence at each timestep, the recent
history constitutes a particular bit-string of length $m$. For general $m$, there will be $P=2^{m}$
possible history bit-strings. These history bit-strings can alternatively be represented in decimal
form: $\mu =\{0,1,...,P-1\}$ where $\mu =0$\ corresponds to $00$, $\mu =1$\ corresponds to 01 etc.  A
strategy consists of a predicted action, $-1$ or $+1$, for each possible history bit-string. Hence
there are
$2^{P}$ possible strategies. For
$m=2$ for example, there are therefore $16$ possible strategies. In order to mimic the
heterogeneity in the system, a typical game setup would have each agent randomly picking $S$
strategies at the outset of the game. In the Minority Game, these strategies are then fixed for all
time -- however a more general setup would allow the strategies held, and hence the heterogeneity in
the population, to change with time. The agents then update the scores of their
strategies after each timestep with
$+1$ (or
$-1$) as the pay-off for predicting the action which won (or lost). This
binary representation of histories and strategies is due to Challet and Zhang \cite{RSS}. 

\subsection{Rules of the Game}
The rules of the game determine the subsequent game dynamics. The particular rules chosen will depend
on the practical system under consideration. It is an open question as to how best to
abstract an appropriate `game' from a real-world Complex System. In Ref. \cite{book}, we discuss
multi-agent games which are relevant to financial markets. Elsewhere we plan to discuss possible
choices of game for the specific case of foraging fungal colonies \cite{mark}. The
foraging mechanism adopted within such networks might well be active in other biological
systems, and may also prove relevant to social/economic networks. Insight into 
network functionality may even prove useful in designing and controlling arrays of imperfect
nanostructures, microchips, nano-bio components, and other `systems on a chip' \cite{damienme}.

The following
rules need to be specified in order to fully define the dynamics of a multi-agent game:
\begin{itemize} 
\item How is the global outcome, and hence the winning action, to be decided at each timestep $t$? In
a general game, the global outcome may be an arbitrary function of the values of
$n_{-1}[t'\leq t]$, $n_{+1}[t'\leq t]$, and $L[t'\leq t]$ for any $t'$.
Consider the specific case of the El Farol Problem: we can define the action
$+1$ ($-1$) to be attend (stay away) with $L[t]$ representing the bar capacity. If
$n_{+1}[t]<L[t]$, the bar will be undercrowded and hence can be assigned the global outcome $0$. Hence
the winning action is
$+1$ (i.e. attend). Likewise if
$n_{+1}[t]>L[t]$, the bar will be overcrowded and can be assigned the global outcome $1$. The winning
action is then $-1$ (i.e. stay away). However more generally, the global outcome and hence winning
action may be any function of present or past system data.  Furthermore, the resource level 
$L[t]$ may be endogenously produced (e.g. a specific function of past values of $n_{-1}[t]$,
$n_{+1}[t]$) or exogenously produced (e.g. determined by external environmental concerns). Figure 1
denotes the global outcome (and hence winning action) algorithm as lying in the hands of a
ficticious `Game-master', who aggregates the information in the game such as $n_{+1}[t]$, $n_{-1}[t]$
and
$L[t]$, and then announces the global outcome. We note that in principle,
the agents themselves do not actually need to know what game they are playing. Instead they are just
fed with the global outcome: each of their $S$ strategies are then rewarded/penalized according to
whether the strategy predicted the winning/losing action. One typical setup has agents
adding/deducting one virtual point from strategies that would have suggested the winning/losing
action.
\item How do agents decide which strategy to use? Typically one might expect agents
to play their highest scoring strategy, as in the original El Farol Problem and Minority Game.
However agents may instead adopt a stochastic method for choosing which strategy to use at each
timestep. There may even be `dumb' agents, who use their worst strategy at each timestep. Of course,
it is not obvious that such agents would then necessarily lose, i.e. whether such a method
is actually dumb will depend on the decisions of the other agents.
\item What happens in a strategy tie-break situation? Suppose agents are programmed to always use
their highest-scoring strategy. If an agent were then to hold two or more strategies that are tied
for the position of highest-scoring strategy, then a rule must be invoked for breaking this tie.
One example would be for the agent to toss a fair coin in order to decide which of his tied
strategies to use at that turn of the game. Alternatively, an agent may prefer to stick with the
strategy which was higher on the previous timestep.
\item What are the rules governing the connections between agents? In terms of structure,
the connections may be hard-wired or temporary, random or ordered or partially ordered (e.g.
scale-free network or small-world network). In terms of functionality, there is also an endless
set of possible choices of rules. For example, any two connected agents might compare the scores of
their highest scoring strategies, and hence adopt the predicted action of whichever strategy is
overall the highest-scoring. Just as above, a tie-break rule will then have to be imposed for the
case where two connected agents have equal-scoring highest strategies -- for example a
coin-toss might be invoked, or a `selfish' rule may be used whereby an agent sticks with his own
strategy in the event of such a tie.  The connections themselves may have directionality, e.g. perhaps
agent
$b$ can influence agent
$d$ but not vice-versa. Or maybe the connection (e.g. between $b$ and $a$) is inactive unless a
certain criterion is met. Hence it may turn out that agent $b$ follows $d$ even though agent
$d$ is actually following agent $a$. 
\item Do agents have to play at each timestep? For simplicity in the present paper, we will consider
that this is the case. This rule is easily generalized by introducing
a confidence level: if the agent doesn't hold any strategies
with sufficiently high success rate, then the agent does not participate at that timestep. This in
turn implies that the number of agents participating at each timestep fluctuates. We do not pursue
this generalization here, but note that its effect is to systematically prevent the playing
of low-scoring strategies which are anticorrelated to the high-scoring strategies. For financial
markets, for example, this extra property of confidence-level is a crucial ingredient for building a
realistic market model, since it leads to fluctuations in the `volume' of active agents
\cite{book,dublin}.
\end{itemize}

\noindent We emphasize that the set-up in Figure 1 does not make any assumptions about the actual
game being played, nor how the winning decision is to be decided. Instead, 
one hopes to obtain a fairly general description of the type of dynamics which are 
possible {\em without} having to specify the exact details of the game. On the other hand, 
situations can arise where the precise details of the strategy tie-break mechanism, for example,
have a fundamental impact on the type of dynamics exhibited by the system and hence
its overall performance in terms of achieving some particular goal. We refer to Ref. \cite{charley}
for an explicit example of this, for the case of network-based multi-agent  systems.

\subsection{Strategy Space}
Figure 2 shows in more detail the $m=2$ example strategy space from Figure 1. 
A strategy is a set of instructions to describe what an agent should do in any given situation, i.e.
given any particular history $\mu$, the strategy then decides what action the agent should take. The
strategy space is the set of strategies from which agents are allocated their strategies. If this
strategy allocation is fixed randomly at the outset, then this acts as a source of quenched disorder.
Alternatively, the strategy allocation may be allowed to evolve in response to the
system's dynamics. In the case that the initial strategy
allocation is fixed, it is clear that the agents playing the game are
limited, and hence may become `frustrated', by this quenched disorder.  The strategy space shown is
known as the Full Strategy Space FSS, and contains all possible permutations of the actions
$-1$ and $+1$ for each history. As such there are $2^{2^{m}}$ strategies in this
space. The $2^{m}$ dimensional hypercube shows all 
$2^{2^{m}}$ strategies from the FSS at its vertices. Of course, there are many additional strategies
that could be thought of, but which aren't present within the FSS. For example, the
simple strategies of persistence and anti-persistence are not present in the FSS. The advantage
however of using the FSS is that the strategies form a complete set, and as such the FSS displays no
bias towards any particular action for a given history. 
To include any additional strategies like
persistence and anti-persistence would mean opening up the strategy space, hence
losing the simplicity of the B-A-R structure and returning to the complexity of Arthur's original
El Farol Problem
\cite{farol,uselfarol}.

It can be observed from the FSS, that one can choose a subset of strategies
\cite{RSS} such that any pair within this subset has one of the following
characteristics:
\begin{itemize}
\item anti-correlated, e.g. $-1-1-1-1$ and $+1+1+1+1$, or $-1-1+1+1$ and $+1+1-1-1$. For example, any
two agents using the ($m=2$) strategies $-1-1+1+1$ and $+1+1-1-1$ respectively, would take the
opposite action irrespective of the sequence of previous outcomes and hence the history. Hence one
agent will always do the opposite of the other agent. For example, if one agent chooses $+1$
at a given timestep, the other agent will choose $-1$. Their net effect on the demand
$D[t]$ therefore cancels out at each timestep, regardless of the history. Hence they will not
contribute to fluctuations in $D[t]$.
\item uncorrelated, e.g. $-1-1-1-1$ and $-1-1+1+1$. For example, any two agents using the 
strategies 
$-1-1-1-1$ and $-1-1+1+1$ respectively, would take the opposite action for two of the four
histories, while they would take the same action for the remaining two histories.
If the $m=2$ histories occur equally often, the actions of the two agents
will be uncorrelated on average.
\end{itemize}
\noindent A convenient measure of the distance (i.e. closeness) of any two strategies
is the Hamming distance which is defined as the number of bits that need to be
changed in going from one strategy to another. For example, the Hamming distance
between $-1-1-1-1$ and $+1+1+1+1$ is $4$, while the Hamming distance between $-1-1-1-1$ and
$-1-1+1+1$ is just $2$. Although there are $2^{P}\equiv 2^{2^{m=2}}\equiv 16$\ strategies
in the
$m=2 $ strategy space, it can be seen that one can choose subsets such that any
strategy-pair within this subset is either anti-correlated or
uncorrelated. Consider, for example, the two groups
\begin{equation} U_{m=2}\equiv \{-1-1-1-1,\ +1+1-1-1,\ +1-1+1-1,\ -1+1+1-1\}
\end{equation} 
and
\begin{equation}
\overline{U_{m=2}}\equiv \{+1+1+1+1,\ -1-1+1+1,\ -1+1-1+1,\ +1-1-1+1\}.
\end{equation} 
Any two strategies within $U_{m=2}$\ are uncorrelated since they
have a Hamming distance of $2$. Likewise any two strategies within
$\overline{U_{m=2}}$\ are uncorrelated since they have a relative Hamming distance of
$2$. However, each strategy in $U_{m=2}$ has an anti-correlated strategy in
$\overline{U_{m=2}}$: for example, $-1-1-1-1$ is anti-correlated to $+1+1+1+1$, $+1+1-1-1$ is
anti-correlated to $-1-1+1+1$ etc. This subset of strategies comprising $U_{m=2}$\ and
$\overline{U_{m=2}}$, forms a Reduced Strategy Space (RSS) \cite{RSS}. Since it
contains the essential correlations of the Full Strategy Space (FSS), running a given game simulation
within the RSS is likely to reproduce the main features obtained using the FSS
\cite{RSS}. The RSS has a smaller number of strategies
$2.2^{m}=2P\equiv 2^{m+1}$ than the FSS which has $2^{P}=2^{2^{m}}$. For $m=2$,
there are $8$ strategies in the RSS compared to $16$ in the FSS, whilst for $m=8$
there are
$1.16\times 10^{77}$ strategies in the FSS but only $512$\ strategies in the RSS.
We note that the choice of the RSS is not unique, i.e. within a given FSS there are
many possible choices for a RSS. In particular, it is possible to create
$2^{2^{m}}/2^{m+1}$ distinct reduced strategy spaces from the FSS. In short, the
RSS provides a minimal set of strategies which `span' the FSS and are hence
representative of its full structure.

\subsection{History Space}
The history $\mu$ of recent outcomes changes in time, i.e. it is a dynamical
variable. The history dynamics can be represented on a
directed graph (a so-called digraph). The particular form of directed graph is
called a de Bruijn graph. Figure 3 shows some examples of the de Bruijn graph for
$m=1,2,$ and $3$. The probability that the outcome at time $t+1$ will be a $1$ (or
$0$) depends on the state at time $t$. Hence it will depend on the previous $m$
outcomes, i.e. it depends on the particular state of the history bit-string. The
dependence on earlier timesteps means that the game is not Markovian. However,
modifying the game such that there is a finite time-horizon for scoring strategies, may then allow
the resulting game to be viewed as a high-dimensional Markov process 
(see Refs. \cite{THMG1,THMG2} for
the case of the Minority Game).

\subsection{Heterogeneity in strategy allocation, and initial conditions}
The dynamics for a particular run of the B-A-R system will depend upon the strategies that
the agents hold, and the random process used to decide tie-breaks. The
particular dynamics which emerge also depend upon the initial score-vector of the
strategies and the initial history used to seed the start of the game. If the initial
strategy score-vector is not `typical', then a bias can be
introduced into the game which never disappears. In short, the system never recovers from this bias.
It will be assumed that no such initial bias exists. In practice this is achieved, for
example, by setting all the initial scores to zero. The initial choice of history is
not considered to be an important effect. It is assumed that any transient effects
resulting from the particular history seed will have disappeared, i.e. the initial
history seed does not introduce any long-term bias. 

The strategy allocation among agents can be described in terms of a tensor
$\Omega
$ \cite{paul}. This tensor $\Omega $ describes the distribution of strategies among
the $N$ individual agents. If this strategy allocation is fixed from the beginning of the game, then
it acts as a quenched disorder in the system. The rank of the tensor $\Omega $\ is given by the
number of strategies $S$ that each agent holds. For example, for $S=3$ the element $\Omega _{i,j,k}$
gives the number of agents assigned strategy $i$, then strategy $j$, and then strategy 
$k$, in that order. Hence
\begin{equation}
\sum_{i,j,k,...}^{X}\Omega _{i,j,k,...}=N,
\end{equation}
where the value of $X$\ represents the number of distinct strategies
that exist within the strategy space chosen: $X=2^{2^{m}}$ in the FSS, and
$X=2.2^{m}$ in the RSS. Figure 4  shows an example distribution $\Omega$ for $N=101$
agents in the case of $m=2$ and $S=2$, in the reduced strategy space RSS. We note that a single
$\Omega$ `macrostate' corresponds to many possible `microstates' describing the specific partitions of
strategies among the agents. For example, consider an $N=2$ agent system with $S=1$: the
microstate $(R,R^\prime)$ in which agent $1$ has strategy $R$ while agent $2$ has strategy
$R^\prime$, belongs to the same macrostate $\Omega$ as $(R^\prime,R)$ in which agent $1$ has strategy
$R^\prime$ while agent $2$ has strategy $R$. Hence the present Crowd-Anticrowd theory retained at 
the level of a given $\Omega$, describes the set of all games which belong to that same $\Omega$
macrostate.  We also note that although $\Omega$ is not symmetric, it can be made so since the agents
will typically not distinguish between the order in which the two strategies are picked. Given this,
we will henceforth focus on $S=2$ and consider the symmetrized version of the strategy allocation
matrix given by
${\Psi}=\frac{1}{2}({\Omega}+{\Omega}^T)$. In general, $\Psi$ would be allowed to change in time,
possibly evolving under some pre-defined selection criteria. 
In David Smith's paper elsewhere in this Workshop, changes in $\Psi$
are invoked in order to control the future evolution of the multi-agent game: this corresponds to a
change in heterogeneity in the agent population, and could represent the physical situation
where individual agents (e.g. robots) could be re-wired, re-programmed, or replaced in order to
invoke a change in the system's future evolution, or to avoid certain future scenarios or
trajectories for the system as a whole.

\subsection{Design Criteria for Collective}
In addition to the excess demand $D[t]$ in such B-A-R systems, one is typically
also interested in higher order-moments of this quantity -- for example, the standard deviation of
$D[t]$ (or `volatility' in a financial context). This gives a measure of the fluctuations in the
system, and hence can be used as a measure of `risk' in the system. In particular the standard
deviation gives an idea of the size of typical fluctuations in the system. However in the case that
$D[t]$ has a power-law distribution, the standard deviation may be a misleading representation of such
risk because of the `heavy tails' associated with the power-law \cite{book}. 
Practical risk is arguably better associated with the probability of the system hitting a certain
critical value ($D_{\rm crit}$), in a similar way to the methodology of financial Value-at-Risk
\cite{book}. However a note of caution is worthwhile: if there are high-order correlations in
the system and hence in the time-series $D[t]$ itself, any risk assessment based on
Probability Distribution Functions over a fixed time-scale (e.g. single timestep) may be
misleading. Instead, it may be the {\em cumulative} effects of a string of large negative
values of $D[t]$ which constitute the true risk, since these may take the system into
dangerous territory. The possibility of designing a Collective in order to build in some form of
immunization/resistance to such large cumulative changes or endogenous `extreme events', or
alternatively the possibility of performing `on-line' soft control in an evolving system, is a
fascinating topic. We again refer to the paper of David Smith elsewhere in this Workshop for a detailed
discussion of these topics. For simplicity, we will focus here on developing a description of
$D[t]$ and its standard deviation, noting that the same analytic approach would work equally well for
other statistical functions of $D[t]$.

\section{Crowd-Anticrowd formalism}
Here we focus on developing a description of $D[t]$ and its standard
deviation. Similar analysis can be carried out for any function of $D[t]$. For example, the function
$f[D[t]]\equiv[D[t]-\langle D[t]\rangle]^\alpha$ with $\alpha>2$, places more weighting on large
deviations of
$D[t]$ from the mean. Alternatively, one may consider a function of past values
$\{D[t'<t]\}$, such as the cumulative value $P[t]=\sum_{i} D[t_i<t]$, or any function of the
constituent processes
$n_{+1}[t]$ and
$n_{-1}[t]$. 

Consider an arbitrary timestep $t$ during a run of the game, at which the particular realization of
the strategy allocation matrix is given by
$\Psi$. We will assume that
$\Psi$ is effectively constant over the time-window during which we will calculate the statistical
properties of $D[t]$. There is a current score-vector
$\underline{S}[t]$ and a current history
$\mu
\lbrack t]$\ which define the state of the game. The excess demand $D[t]=D\left[ \underline{S}[t],\mu
\lbrack t]\right]$ is given by Equation 1. 
The standard deviation of $D[t]$ for this given run, corresponds
to a time-average for a given realization of 
$\Psi
$ and a given set of initial conditions. It may turn out that we are only interested in
ensemble-averaged quantities: consequently the standard deviation will then need to be averaged
over many runs, and hence averaged over all realizations of 
$\Psi $ \emph{and} all sets of initial conditions. 

Equation 1 can be rewritten by summing over the RSS as
follows:
\begin{equation} D\left[ \underline{S}[t],\mu \lbrack t]\right]
=n_{+1}[t]-n_{-1}[t]\equiv \sum_{R=1}^{2P}a_{R}^{\mu
\lbrack t]}n_{R}^{\underline{S}[t]},
\end{equation}
where $P=2^{m}$. The quantity $a_{R}^{\mu \lbrack t]}=\pm 1$ is the
response of strategy $R$ to the history bit-string $\mu $\ at time $t$. The quantity
$n_{R}^{\underline{S}[t]}$ is the number of agents using strategy $R$ at time $t$. The superscript
$\underline{S}[t]$ is a reminder that this number of agents will depend on the strategy score at time
$t$. The calculation of the average demand $D[t]$ will now be shown, where the average is
over time for a given realization of the strategy allocation $\Psi $. We use the
notation $\left\langle X[t]\right\rangle _{t}$ to denote a time-average over the variable $X[t]$
for a given $\Psi $. Hence
\begin{eqnarray}
\left\langle D\left[ \underline{S}[t],\mu \lbrack t]\right] \right\rangle _{t}
&=&\sum_{R=1}^{2P}\left\langle a_{R}^{\mu \lbrack
t]}n_{R}^{\underline{S}[t]}\right\rangle _{t} \\ &=&\sum_{R=1}^{2P}\left\langle
a_{R}^{\mu \lbrack t]}\right\rangle _{t}\left\langle
n_{R}^{\underline{S}[t]}\right\rangle _{t}  \nonumber
\end{eqnarray}
where we have used the property that $a_{R}^{\mu \lbrack t]}$ and
$n_{R}^{\underline{S}[t]}$ are uncorrelated.
We now consider the special case in which all
histories are visited equally on average: this may arise as the result of a periodic cycling through
the history space (e.g. a Eulerian trail around the de Bruijn graph) or if the histories are visited
randomly. Even if this situation does not hold for a specific $\Psi$, it may indeed hold once the
averaging over 
$\Psi$ has also been taken. For example, in the Minority Game all histories are visited
equally at small
$m$ and a given $\Psi$: however 
the same is only true for large $m$  
if  we take the additional average over all $\Psi$.  
Under the property of equal histories, we can write 
\begin{eqnarray}
\left\langle D\left[ \underline{S}[t],\mu \lbrack t]\right] \right\rangle _{t}
&=&\sum_{R=1}^{2P}\left( \frac{1}{P}\sum_{\mu =0}^{P-1}a_{R}^{\mu \lbrack
t]}\right) \left\langle n_{R}^{\underline{S}[t]}\right\rangle _{t}  \label{hist} \\
&=&\sum_{R=1}^{P}\left( \frac{1}{P}\sum_{\mu =0}^{P-1}a_{R}^{\mu \lbrack
t]}+a_{\overline R}^{\mu \lbrack
t]}\right) \left\langle n_{R}^{\underline{S}[t]}\right\rangle _{t}  \nonumber \\
&=&\sum_{R=1}^{P}0.\left\langle n_{R}^{\underline{S}[t]}\right\rangle _{t} \nonumber
\\ &=&0  \nonumber 
\end{eqnarray}
where we have used the exact result that $a_{R}^{\mu \lbrack t]}=-a_{\overline R}^{\mu \lbrack t]}$
for all
$\mu[t]$, and the approximation  
$\left\langle n_{R}^{\underline{S}[t]}\right\rangle _{t}=\left\langle
n_{\overline R}^{\underline{S}[t]}\right\rangle _{t}$. This approximation is reasonable for a
competitive game since there is typically no a priori best strategy, hence each strategy has its
own `five minutes of glory': if the strategies are distributed fairly evenly among the agents,
this then implies that the average number playing each strategy is approximately equal and
hence $\left\langle n_{R}^{\underline{S}[t]}\right\rangle _{t}=\left\langle
n_{\overline R}^{\underline{S}[t]}\right\rangle _{t}$.  In the event that all 
histories are not equally visited over time, even after averaging over all $\Psi$, it may still
happen that the system's dynamics is restricted to equal visits to some {\em subset} of histories. An
example of this would arise for
$m=3$, for example, for a repetitive sequence of outcomes $\dots 00110011001$ in which case the
system repeatedly performs the 4-cycle $\dots\rightarrow 001\rightarrow 011\rightarrow 110\rightarrow
100\rightarrow 001\rightarrow\dots$. In this case one can then carry out the averaging in
Equation \ref{hist} over this subspace of four histories, implying that there are now strategies that
are effectively identical (i.e. they have the same response for these four histories, even though they
differ in their response for one or more of the remaining four histories $000$, $010$, $101$, $111$
which are not visited). More generally, such sub-cycles within the de Bruijn graph may lead to a bias
towards
$1$'s or
$0$'s in the global outcome sequence. We note that such a biased series of outcomes can also be
generated by biasing the initial strategy pool. 

We will focus now on the fluctuations of
$D[t]$ about its average value. The
variance of $D[t]$, which is the square of the standard deviation, is given by
\begin{equation}
\sigma _{\Psi}^{2} =\left\langle D\left[ \underline{S}[t],\mu \lbrack t]
\right] ^{2}\right\rangle _{t}-\left\langle D\left[ \underline{S}[t],\mu
\lbrack t]\right] \right\rangle _{t}^{2}  \ .
\end{equation}
For simplicity, we will assume the game output is unbiased and hence we can set $\left\langle
D\left[
\underline{S}[t],\mu
\lbrack t]\right] \right\rangle _{t}=0$. As mentioned above, this might arise in a given system for
a given $\Psi$, or after additional averaging over all possible configurations $\{\Psi\}$.  We stress
that this approximation does not {\em have} to be made -- one can simply continue subtracting
$\left\langle D\left[
\underline{S}[t],\mu
\lbrack t]\right] \right\rangle _{t}^2$ from the right hand side of the expression for
$\sigma_{\Psi}^2$. Hence
\begin{eqnarray}
\sigma _{\Psi}^{2} &=&\left\langle D\left[
\underline{S}[t],\mu \lbrack t]\right] ^{2}\right\rangle _{t}  \\
&=&\sum_{R,R^{\prime }=1}^{2P}\left\langle a_{R}^{\mu \lbrack
t]}n_{R}^{\underline{S}[t]}a_{R^{\prime }}^{\mu \lbrack t]}n_{R^{\prime
}}^{\underline{S}[t]}\right\rangle _{t}.  \nonumber
\end{eqnarray}
In the case that the system visits all possible histories equally, the double sum can usefully be
broken down into three parts, based on the correlations between the strategies:
\underline{$a_{R}$}$.
\underline{a_{R^{\prime }}}=P$\ (fully correlated), \underline{$a_{R}$}$.
\underline{a_{R^{\prime }}}=-P$\ (fully anti-correlated), and \underline{$
a_{R}$}$.\underline{a_{R^{\prime }}}=0$ (fully uncorrelated) where 
$\underline{a_{R}}$ is a vector of dimension $P$ with $R$'th component $a_{R}^{\mu \lbrack t]}$.
This decomposition is exact in the RSS in which we are working. Again we note that if all histories
are not equally visited, yet some subset are, then this averaging can be carried out over the
restricted subspace of histories. The equal-histories case yields
\begin{eqnarray}
\sigma _{\Psi}^{2} &=&\sum_{R=1}^{2P}\left\langle \left( a_{R}^{\mu
\lbrack t]}\right) ^{2}\left( n_{R}^{\underline{S}[t]}\right) ^{2}\right\rangle
_{t}+\sum_{R=1}^{2P}\left\langle a_{R}^{\mu \lbrack t]}a_{
\overline{R}}^{\mu \lbrack t]}n_{R}^{\underline{S}[t]}n_{\overline{R}}^{
\underline{S}[t]}\right\rangle _{t}+
\sum_{R\neq R^{\prime }\neq \overline{R}}^{2P}\left\langle a_{R}^{\mu \lbrack
t]}a_{R^{\prime }}^{\mu \lbrack t]}n_{R}^{\underline{S}[t]}n_{R^{\prime
}}^{\underline{S}[t]}\right\rangle _{t}  \nonumber \\ &=&\sum_{R=1}^{2P}\left\langle
\left( n_{R}^{\underline{S}[t]}\right)
^{2}-n_{R}^{\underline{S}[t]}n_{\overline{R}}^{\underline{S}[t]}\right\rangle
_{t}+\sum_{R\neq R^{\prime }\neq \overline{R}}^{2P}\left\langle a_{R}^{\mu \lbrack
t]}a_{R^{\prime }}^{\mu \lbrack t]}\right\rangle _{t}\left\langle
n_{R}^{\underline{S}[t]}n_{R^{\prime }}^{
\underline{S}[t]}\right\rangle _{t} \\ &=&\sum_{R=1}^{2P}\left\langle \left(
n_{R}^{\underline{S}[t]}\right)
^{2}-n_{R}^{\underline{S}[t]}n_{\overline{R}}^{\underline{S}[t]}\right\rangle
_{t}+\sum_{R\neq R^{\prime }\neq \overline{R}}^{2P}\left( 
\frac{1}{P}\sum_{\mu =0}^{P-1}a_{R}^{\mu \lbrack t]}a_{R^{\prime }}^{\mu
\lbrack t]}\right) \left\langle n_{R}^{\underline{S}[t]}n_{R^{\prime
}}^{\underline{S}[t]}\right\rangle _{t}  \nonumber \\ &=&\sum_{R=1}^{2P}\left\langle
\left( n_{R}^{\underline{S}[t]}\right)
^{2}-n_{R}^{\underline{S}[t]}n_{\overline{R}}^{\underline{S}[t]}\right\rangle
_{t}.  \nonumber
\end{eqnarray}
This sum over $2P$ terms can be written equivalently as a sum over $P$
terms,
\begin{eqnarray}
\sigma _{\Psi}^{2} &=&\sum_{R=1}^{2P}\left\langle \left(
n_{R}^{\underline{S}[t]}\right)
^{2}-n_{R}^{\underline{S}[t]}n_{\overline{R}}^{\underline{S}[t]}\right\rangle _{t} 
\label{requ} \\ &=&\sum_{R=1}^{P}\left\langle
\left( n_{R}^{\underline{S}[t]}\right)
^{2}-n_{R}^{\underline{S}[t]}n_{\overline{R}}^{\underline{S}[t]}+\left(
n_{\overline{R}}^{\underline{S}[t]}\right)
^{2}-n_{\overline{R}}^{\underline{S}[t]}n_{R}^{\underline{S}[t]}\right\rangle _{t} 
\nonumber \\ &=&\sum_{R=1}^{P}\left\langle \left( n_{R}^{\underline{S}[t]}\right)
^{2}-2n_{R}^{\underline{S}[t]}n_{\overline{R}}^{\underline{S}[t]}+\left(
n_{\overline{R}}^{\underline{S}[t]}\right) ^{2}\right\rangle _{t}  \nonumber \\
&=&\sum_{R=1}^{P}\left\langle \left(
n_{R}^{\underline{S}[t]}-n_{\overline{R}}^{\underline{S}[t]}\right)
^{2}\right\rangle _{t} \equiv \left\langle \sum_{R=1}^{P}\left(
n_{R}^{\underline{S}[t]}-n_{\overline{R}}^{\underline{S}[t]}\right)
^{2}\right\rangle _{t}.  \nonumber
\end{eqnarray}
The values of $n_{R}^{\underline{S}[t]}$ and
$n_{\overline{R}}^{\underline{S}[t]}$\ for each $R$ will depend on the precise form
of $\Psi $. 
We now proceed to consider the ensemble-average 
over all possible realizations of the strategy allocation matrix $\Psi$. 
The ensemble-average is denoted as $\left\langle ...\right\rangle _{\Psi}$, and
for simplicity the notation $\left\langle \sigma _{\Psi}^{2}\right\rangle
_{\Psi}=\sigma ^{2}$ is defined. This ensemble-average is performed on either
side of Equation \ref{requ},
\begin{equation}
\sigma ^{2}=\left\langle \left\langle \sum_{R=1}^{P}\left(
n_{R}^{\underline{S}[t]}-n_{\overline{R}}^{\underline{S}[t]}\right)
^{2}\right\rangle _{t}\right\rangle _{\Psi } \label{ModelC1}
\end{equation}
yielding the variance in the excess demand $D[t]$.
Equation \ref{ModelC1} is an important intermediary result for the
Crowd-Anticrowd theory. It is straightforward to obtain analogous expressions for the variances in
$n_{+1}[t]$ and
$n_{-1}[t]$. 

\subsection{Quantitative theory}
Equation \ref{ModelC1} provides us with
an expression for the time-averaged fluctuations. 
Some form of approximation must be introduced in order to reduce Equation \ref{ModelC1} to
explicit analytic expressions. It turns out that Equation \ref{ModelC1} can be manipulated in a
variety of ways, depending on the level of approximation that one is prepared to make. The precise
form of any resulting analytic expression will depend on the details of the approximations
made.

We now turn to the problem of evaluating Equation \ref{ModelC1}
analytically. A key first step is to relabel the strategies. Specifically, the sum in Equation
\ref{ModelC1} is re-written to be over a \emph{virtual-point ranking} $K$\ and not
the decimal form
$R$. Consider the variation in points for a given strategy, as a function of time for
a given realization of $\Psi $. The ranking (i.e. label) of a given
strategy in terms of virtual-points score will typically change in time since the
individual strategies have a variation in virtual-points which also varies in time. For the
Minority Game, this variation is quite rapid in the low $m$ regime since there are many more
agents than available strategies -- hence any strategy emerging as the instantaneously
highest-scoring, will immediately get played by many agents and therefore be likely to lose on the
next time-step. More general games involving competition within a multi-agent population, will
typically generate a similar ecology of strategy-scores with no all-time winner. [N.B. If this
weren't the case, then there would by definition be an absolute best, second-best
etc. strategy. Hence any agent observing the game from the outside would be able to choose such a
strategy and win consistently. Such a system is not sufficiently
competitive, and is hence not the type of system we have in mind]. This implies
that the specific identity of the `$K$'th highest-scoring strategy' changes frequently in time. It
also implies that
$n_{R}^{\underline{S}[t]}$ varies considerably in time. Therefore in order to proceed, 
we shift the focus onto the time-evolution of the highest-scoring
strategy, second highest-scoring strategy etc. This should have a much smoother time-evolution than
the time-evolution for a given strategy. In short, the focus is shifted from the time-evolution of the
virtual points of a given strategy (i.e. from $S_{R}[t]$) to the time-evolution of
the virtual points of the $K$'th highest scoring strategy (i.e. to $S_{K}[t]$).

Figure 5 shows a schematic representation of how the scores of the two top
scoring strategies might vary, using the new virtual-point ranking scheme. Also shown are the
lowest-scoring two strategies, which at every timestep are obviously just the anticorrelated
partners of the instantaneously highest-scoring two strategies. In the case that the strategies all
start off with zero points, these anticorrelated strategies appear as the mirror-image, i.e.
$S_{K}[t]=-S_{\overline K}[t]$.  The label
$K$ is used to denote the rank in terms of strategy score, i.e.
$K=1$ is the highest scoring strategy position, $K=2$ is the second highest-scoring strategy
position etc. with
\begin{equation} S_{K=1}>S_{K=2}>S_{K=3}>S_{K=4}>...
\end{equation}
assuming no strategy-ties. (Whenever strategy ties occur, this ranking gains a `degeneracy' in that
$S_{K}=S_{K+1}$ for a given $K$). A
given strategy, e.g.
$-1-1-1-1$, may at a given timestep have label
$K=1$, while a few timesteps later have label $K=5$. Given that
$S_{R}=-S_{\overline{R}}$\ (i.e. all strategy scores start off at zero), then we
know that
$S_{K}=-S_{\overline{K}}$. Equation \ref{ModelC1} can hence be rewritten exactly
as
\begin{equation}
\sigma^{2}=\left\langle \left\langle \sum_{K=1}^{P}\left(
n_{K}^{\underline{S}[t]}-n_{\overline{K}}^{\underline{S}[t]}\right)
^{2}\right\rangle _{t}\right\rangle _{\Psi }.  \label{ModelC1K}
\end{equation} 
Now we make another important observation. Since in the systems of interest the agents are
typically playing their highest-scoring strategies, then the relevant quantity in determining
how many agents will instantanously play a given strategy, is a knowledge of its relative ranking --
not the actual value of its virtual points score. This suggests that the quantities
$n_{K}^{\underline{S}[t]}$\ and
$n_{\overline{K}}^{\underline{S}[t]}$\ will fluctuate relatively little in time, and that
we should now develop the problem in terms of time-averaged values.

The actual number of agents $n_{K}^{\underline{S}[t]}$ playing the $K$'th
ranked strategy at timestep $t$, can be determined from knowledge of the strategy allocation matrix
$\Psi$ and the strategy scores ${S}[t]$, in order to calculate how many agents hold
the $K$'th ranked strategy but {\em do not} hold another strategy with higher-ranking. The
heterogeneity in the population represented by
$\Psi$, combined with the strategy scores ${S}[t]$, determines
$n_{K}^{\underline{S}[t]}$ for each $K$ and hence the standard deviation in $D[t]$.  
We can rewrite the number of agents playing the strategy in position $K$ at any
timestep $t$, in terms of some constant value $n_K$ plus a fluctuating term :
\begin{equation} n_{K}^{\underline{S}[t]}=n_{K}+\varepsilon _{K}[t]\ .
\end{equation}
Consider a given timestep $t$ in the game's evolution. Some strategies may be tied in virtual-points
(i.e. they are `degenerate') while others are not (i.e. `non-degenerate'). Depending on the rules
of the game, such degeneracy may imply one of the following: (i) an agent must throw a
coin to break a tie between any two degenerate  strategies in
his possession. In this case, any two agents holding the same pair of degenerate strategies may then
disagree as to the ranking of the two strategies following the tie-break coin toss. (ii) The
Game-master must throw a master coin in order to produce a non-degenerate ranking. In this case, all
agents will then agree on a unique virtual-point ranking of strategies. (iii) An agent must preserve
the ranking from the most recent timestep at which the two strategies weren't degenerate.
Depending on the exact microscopic rule adopted, the global dynamics may differ. We will generally
assume that rule (i) holds, since rule (ii) might introduce large random fluctuations while rule
(iii) could favor deterministic cyclic behavior.
Suppose that an agent holds two strategies $R$ and $R'$, and that they are non-degenerate
at timestep $t$. Suppose that they occupy ranks $K$ and $K+1$ respectively. Let the number of agents 
playing strategy $R$, the $K$'th ranked strategy,  be $n_K$ at timestep $t$. Similarly the 
number of agents 
playing strategy $R'$, the $(K+1)$'th ranked strategy,  is $n_{K+1}$ at timestep $t$.
Hence $\varepsilon _{K}[t]=0$ and $\varepsilon _{K+1}[t]=0$ for this timestep, and indeed for all
subsequent $t$ until the next strategy-tie. Suppose that the two strategies $R$ and $R'$ now become
degenerate at timestep $t+1$. In a game where each agent resolves strategy tie-breaks using a
coin-toss, then both
$R$ and
$R'$ have probability $1/2$ of being assigned the rank $K$ at this timestep. This means that any
agent holding strategies $R$ and $R'$, will play $R$ or $R'$ with probability $1/2$. Hence 
such degeneracy acts to add disorder into the number of agents playing the $K$'th highest-scoring
strategy. In particular, such coin-tosses at a given timestep $t'$ will tend to reduce the number of
agents playing higher-scoring strategies at that timestep (i.e. $\varepsilon_K[t']<0$ for small $K$)
but will increase the number of agents
playing lower-scoring strategies (i.e. $\varepsilon_K[t']>0$ for large $K$). 
Therefore if we were to completely ignore coin-tosses, we would tend to
overestimate the number of people playing strategies which are higher-ranking (i.e. small $K$) and
underestimate the number playing strategies which are lower-ranking (i.e. large $K$). This means that
we will overestimate the size of Crowds, and underestimate the size of Anticrowds. Hence we will {\em
overestimate} the value of the standard deviation of demand. [For this reason, the analytic
expression $\sigma^{{\rm delta}\ f}$ that we will derive, slightly {\em overestimates} the actual
value, and hence can be regarded as an approximate {\em upper}-bound as shown later in Figure 7].
Accounting for the correct fraction of degenerate vs. non-degenerate timesteps, will yield a more
accurate calculation of the time-average $\langle \dots
\rangle_t$ in Equation 14. We will adopt this latter approach elsewhere for the network B-A-R system
\cite{charley}, thereby obtaining accurate analytic expressions which are in excellent quantitative
agreement with numerical simulations.

Since our focus in this paper is more general, let us for the moment assume that
we can choose a suitable constant $n_K$ such that the fluctuation $\varepsilon _{K}[t]$
represents a small noise term. Hence,
\begin{eqnarray}
\sigma^{2} &=&\left\langle \sum_{K=1}^{P}\left\langle \left[
n_{K}+\varepsilon _{K}[t]-n_{\overline{K}}-\varepsilon _{\overline{K}}[t]
\right] ^{2}\right\rangle _{t}\right\rangle _{\Psi } \label{average}\\ &=&\left\langle
\sum_{K=1}^{P}\left\langle \left[ (n_{K}-n_{\overline{K}})+(\varepsilon
_{K}[t]-\varepsilon _{\overline{K}}[t])\right] ^{2}\right\rangle _{t}\right\rangle
_{\Psi }  \nonumber \\ &=&\left\langle
\sum_{K=1}^{P}\left\langle \left[ n_{K}-n_{\overline{K}}\right] ^{2}+\left[
\varepsilon _{K}[t]-\varepsilon _{\overline{K}}[t]\right] ^{2}+\left[
2(n_{K}-n_{\overline{K}})(\varepsilon _{K}[t]-\varepsilon _{\overline{K}}[t])\right]
\right\rangle _{t}\right\rangle _{\Psi }  \nonumber \\ &\approx&\left\langle
\sum_{K=1}^{P}\left\langle \left[ n_{K}-n_{\overline{K}}\right] ^{2}\right\rangle _{t}\right\rangle
_{\Psi }=
\left\langle \sum_{K=1}^{P}\left[ n_{K}-n_{\overline{K}}\right]
^{2}\right\rangle _{\Psi },  \nonumber
\end{eqnarray}
since the latter two terms involving noise will average out to be
small. The
resulting expression in Equation
\ref{average} involves no time dependence. The averaging over
$\Psi$ can then be taken inside the sum. The individual terms in the sum, i.e.
$\left\langle \left[ n_{K}-n_{\overline{K}}\right] ^{2}\right\rangle _{\Psi}$,
are just an expectation value of a function of two variables $n_{K}$ and
$n_{\overline{K}}$. Each term can therefore be rewritten exactly using the joint
probability distribution for $n_{K}$ and $n_{\overline{K}}$, which we shall call
$P(n_{K},n_{\overline{K}})$. Hence
\begin{eqnarray}
\sigma^{2} &=&\sum_{K=1}^{P}\left\langle \left[
n_{K}-n_{\overline{K}}\right] ^{2}\right\rangle _{\Psi}  \label{general} \\
&=&\sum_{K=1}^{P}\sum_{n_{K}=0}^{N}\sum_{n_{\overline{K}}=0}^{N}
\left[ n_{K}-n_{\overline{K}}\right] ^{2}P(n_{K},n_{\overline{K}}),  \nonumber
\end{eqnarray}
where the standard probability result involving functions of two
variables has been used.

In the event that Equation \ref{general} is a reasonable approximation for the system under study, 
the question then arises as to how to evaluate it. In general, its value will depend on the detailed
form of the joint probability function
$P(n_{K},n_{\overline{K}})$ which in turn will depend on the ensemble of quenched
disorders $\{\Psi\}$ which are being averaged over. 
We start off by looking at Equation \ref{general} in the limiting case where the averaging
over the quenched disorder matrix is dominated by matrices $\Psi$ which are
nearly flat. This will be a good approximation in the `crowded' limit of small $m$ in which there are
many more agents than available strategies, since
the standard deviation of an element in $\Psi$ (i.e. the standard deviation in bin-size)
is then much smaller than the mean bin-size. [N.B. If $\Omega$ is approximately flat, then so is
$\Psi$]. In this limiting case, there are
several nice features:
\begin{itemize}
\item Given the ranking in terms of virtual-points, i.e.
$S_{K=1}>S_{K=2}>S_{K=3}>S_{K=4}>...$ which holds by definition of the labels 
$\{K\}$ if we neglect tie-breaks, we will also have
\begin{equation} n_{K=1}>n_{K=2}>n_{K=3}>n_{K=4}>...  \nonumber
\end{equation} 
Hence the rankings in terms of highest virtual-points and popularity are identical.
By contrast, the ordering in terms of the labels $\{R\}$ would not be
sequential, i.e. it is \emph{not} true that $n_{R=1}>n_{R=2}>n_{R=3}>n_{R=4}>...$.
\item The strategy $\overline{K}$, which is anticorrelated to
strategy $K$, occupies position $\overline{K}=2P+1-K$\ in this popularity-ranked
list.
\item The probability distribution $P(n_{K},n_{\overline{K}})$ will be sharply
peaked around the $n_{K}$ and $n_{\overline{K}}$ values given by the mean values
for a flat quenched-disorder matrix $\Psi $. We will label these mean values as
${{\overline{n_{K}}}}$ and ${{\overline{n_{\overline{K}}}}}$.
\end{itemize}

\bigskip

\noindent The last point implies that $P(n_{K},n_{\overline{K}})=
\delta_{n_{K},{{\overline{n_{K}}}}}\delta_{n_{\overline{K}},{\overline {n_{\overline{K}}}}}$
and so
\begin{equation}
\sigma^{2}=\sum_{K=1}^{P}\left[
{\overline{n_{K}}}-{{\overline {n_{\overline{K}}}}}\right] ^{2}.  \label{final}
\end{equation} 
We note that there is a very simple interpretation of  Equation
\protect\ref{final}. It represents the sum of the variances for each Crowd-Anticrowd
pair. For a given strategy $K$ there is an anticorrelated strategy $\overline{K}$.
The ${\overline{n_{K}}}$ agents using strategy $K$\ are doing the
\emph{opposite} of the ${{\overline {n_{\overline{K}}}}}$\ agents using strategy
$\overline{K}$\
\emph{irrespective} of the history bit-string. Hence the effective group-size for each
Crowd-Anticrowd pair is
$n_K^{eff}={\overline{n_{K}}}-{{\overline {n_{\overline{K}}}}}\ $ : this represents the net step-size
$d$ of the Crowd-Anticrowd pair in a random-walk contribution to the total variance. Hence, the net
contribution by this Crowd-Anticrowd pair to the variance is given by
\begin{equation}
[\sigma^{2}]_{K\overline{K}} = 4pqd^{2} = 4.\frac{1}{2}.\frac{1}{2}[n_K^{eff}]^{2}= 
\left[{\overline{n_{K}}}-{{\overline {n_{\overline{K}}}}}\right]^{2}
\end{equation}
where $p=q=1/2$ for a random walk. Since all the strong correlations have been
removed (i.e. anti-correlations) it can therefore be assumed that the separate
Crowd-Anticrowd pairs execute random walks which are \emph{uncorrelated} with
respect to each other. [Recall the properties of the RSS - all the remaining
strategies are uncorrelated.] Hence the total variance is given by the sum of the
individual variances,
\begin{equation}
\sigma^{2}=\sum_{K=1}^{P}[\sigma^{2}]_{K\overline{K}}=\sum_{K=1}^{P}\left[
{\overline{n_{K}}}-{{\overline {n_{\overline{K}}}}}\right] ^{2},
\end{equation}
which corresponds exactly to Equation \ref{final}. If strategy-ties occur frequently,
then one has to be more careful about evaluating ${\overline{n_{K}}}$ since its value may be affected
by the tie-breaking rule. We will show elsewhere that this becomes quite important in the case of very
small
$m$ in the presence of network connections - this is because very small $m$ naturally leads to
crowding in strategy space and hence mean-reverting virtual scores for strategies. This
mean-reversion is amplified further by the presence of network connections which increases the
crowding, thereby increasing the chance of strategy ties.

\section{Implementation of Crowd-Anticrowd theory} 
In this section we will consider the application of the
Crowd-Anticrowd theory in some limiting cases. As suggested by the previous
discussion, we will break the implementation of the Crowd-Anticrowd theory down into two separate
regimes: small
$m$, corresponding to many more agents than available strategies, and large $m$ corresponding to the
opposite case. Hence these two regimes are defined by the ratio of the number of
strategies to agents being much less/greater than unity, and hence the strategy allocation matrix
$\Psi$ being densely/sparsely filled.
 
\subsection{Flat quenched disorder matrix $\Psi$, small $m$}
Each element of $\Psi$ has a mean of
$N/(2P)^{S}$ agents per `bin'. In the case of small $m$ and hence densely-filled $\Psi$, the
fluctuations in the number of agents per bin will be small compared to this mean value.  For the case
$S=2$, the mean number of agents whose highest scoring strategy is the strategy occupying
position $K$ at timestep
$t$, will therefore be given by summing the appropriate rows and columns of this
quenched disorder matrix $\Psi$. Figure 6 provides a schematic
representation of $\Psi$ with $m=2$, $s=2$, in the RSS. If the matrix $\Psi$ is flat, then any 
re-ordering due to changes in the strategy ranking
has no effect on the form of the matrix. Therefore the number
of agents playing the $K$'th highest-scoring strategy, will always be proportional to the number
of shaded bins at that $K$ (see Fig. 6 for $K=3$). The shaded elements in Figure 6 therefore represent
those agents who hold a strategy that is ranked third highest in score, i.e. $K=3$. 
In games where the agents use their highest
scoring strategy, any agent
using the strategy in position $K=3$ cannot have any strategy with a higher
position. Hence the agents using the strategy in position $K=3$ must lie
in one of the shaded bins. Since it is assumed that the coverage of the bins is
uniform, the mean number of agents using the strategy in position $K=3$ is given
by
\begin{eqnarray}
{{\overline {n_{K=3}}}} &=&N.\frac{1}{(2P)^{2}}\sum (shaded\text{ }bins) \\
&=&N.\frac{1}{64}.[(8-3)+(8-3)+1]  \nonumber \\ &=&\frac{11}{64}N.  \nonumber
\end{eqnarray}
For more general $m$ and $s$ values this becomes
\begin{eqnarray}
{\overline{n_{K}}} &=&\frac{N}{(2P)^{S}}[S(2P-K)^{S-1}+\frac{S(S-1)}{2}
(2P-K)^{S-2}+...+1]  \label{YofR} \\
&=&\frac{N}{(2P)^{S}}\sum_{r=0}^{S-1}\frac{S!}{(S-r)!r!}[2P-K]^{r}  \nonumber \\
&=&\frac{N}{(2P)^{S}}([2P-K+1]^{S}-[2P-K]^{S})  \nonumber \\ &=&N.\left( \left[
1-\frac{(K-1)}{2P}\right] ^{S}-\left[ 1-\frac{K}{2P}
\right] ^{S}\right) ,  \nonumber
\end{eqnarray}
with $P\equiv 2^{m}$. In the case where each agent
holds two strategies, $S=2,$ ${\overline{n_{K}}}$ can be simplified to
\begin{equation}
{\overline{n_{K}}} = N.\left( \left[ 1-\frac{(K-1)}{2P}\right] ^{2}-\left[ 1-
\frac{K}{2P}\right] ^{2}\right)  
= \frac{(2^{m+2}-2K+1)}{2^{2(m+1)}}N\ .  \label{k} 
\end{equation}
Similarly for ${{\overline{n_{\overline{K}}}}}$ the simplification is as
follows:
\begin{equation}
{{\overline {n_{\overline{K}}}}} = \frac{(2^{m+2}-2\overline{K}+1)}{2^{2(m+1)}}N
= \frac{(2K-1)}{2^{2(m+1)}}N,  \label{kbar} 
\end{equation}
where the relation $\overline{K}=2P-K+1\equiv 2^{m+1}-K+1$\ has been used.
It is emphasized that these results depend on the assumption that the averages are
dominated by the effects of flat distributions of the quenched disorder matrix
$\Psi$, and hence will only be quantitatively valid for low $m$.
Using Equations \ref{k}\ and \ref{kbar}\ in Equation \ref{final} gives
\begin{eqnarray}
\sigma^{2} &=&\sum_{K=1}^{P}\left[ \frac{(2^{m+2}-2K+1)}{
2^{2(m+1)}}N-\frac{(2K-1)}{2^{2(m+1)}}N\right] ^{2} \\
&=&\frac{N^{2}}{2^{2(2m+1)}}\sum_{K=1}^{P}[2^{m+1}-2K+1]^{2}  \nonumber \\
&=&\frac{N^{2}}{3.2^{m}}(1-2^{-2(m+1)}),  \nonumber
\end{eqnarray}
and hence
\begin{equation}
\sigma^{{\rm delta}\  f}=\frac{N}{\sqrt{3}.2^{m/2}}
(1-2^{-2(m+1)})^{\frac{1}{2}},  \label{flatom}
\end{equation} which is valid for small $m$. (The rationale behind the choice of superscript `${\rm
delta}\ f$' will become apparent shortly.)
This derivation has assumed that there are no strategy ties -- more precisely, we have assumed that
the game rules governing strategy ties do not upset the identical forms of the rankings in terms of
highest virtual points and popularity. As discussed earlier, this tends
to overestimate the size of the Crowds using high-ranking strategies, and underestimate the size of
the Anticrowds using low-ranking strategies. Hence the Crowd-Anticrowd cancellation is
underestimated, and consequently $\sigma^{{\rm delta}\  f}$ will overestimate the actual $\sigma$
value. As we will see later in Figure 7, $\sigma^{{\rm delta}\  f}$ does indeed act as an approximate
upper bound.

\subsection{Non-flat quenched disorder matrix $\Psi$, small $m$}
The appearance of a significant number of non-flat quenched disorder matrices
$\Psi$ in the ensemble, implies that the standard deviation for each `bin' is now
significant, i.e. non-negligible compared to the mean. This will be increasingly
true as $m$ increases. In this case, the general analysis is much more complicated,
and should really appeal to the dynamics. However, an approximate theory can still be developed. The
features for the case of ensembles containing a significant number of non-flat quenched disorder
matrices $\Psi$ become as follows:
\begin{itemize}
\item  By definition of the labels $\{K\}$, the ranking in terms of virtual-points
is retained, i.e. S$_{K=1}>$S$_{K=2}>$S$_{K=3}>$S$_{K=4}>...$ is still always true in the absence of
strategy ties. However the disorder in the matrix $\Psi$
may distort the number of agents playing a given strategy to such an extent, that it is no longer
true that 
$n_{K=1}>n_{K=2}>n_{K=3}>n_{K=4}>...$. Hence the rankings in terms of highest virtual-points and
popularity are no longer identical. Another way of saying this is that the disorder in $\Psi$ has
introduced a disorder into the popularity ranking, such that it now differs from the virtual-point
ranking. [We note that this disordering effect in the popularity ranking arising from the non-flat
$\Psi$, occurs in addition to the disordering effect due to coin-toss tie-breaking of
strategy degeneracies discussed earlier].
\item In general we now have that $n_{K^{^{\prime }}}>n_{K^{^{\prime \prime
}}}>n_{K^{^{\prime
\prime \prime }}}>n_{K^{^{\prime \prime \prime \prime }}}>...$, where the label
$K^{\prime }$ need not equal $1$, and $K^{\prime \prime }$ need not equal $2$ etc..
It is however possible to introduce a new label $\{Q\}$ which will rank the
strategies in terms of popularity, i.e.
\begin{equation} n_{Q=1}>n_{Q=2}>n_{Q=3}>n_{Q=4}>...,
\end{equation} where $Q=1$ represents $K^{\prime }$, $Q=2$ represents $K^{\prime
\prime }$, etc.
\end{itemize} 
\noindent With this in mind, we will return to the original general form for the
standard deviation of the excess demand  in Equation \ref{general}, but rewrite it slightly as
follows:
\begin{eqnarray}
\sigma^{2} &=&\sum_{K=1}^{P}\sum_{n_{K}=0}^{N}\sum_{n_{
\overline{K}}=0}^{N}\left[ n_{K}-n_{\overline{K}}\right] ^{2}P(n_{K},n_{
\overline{K}}) \\
&=&\frac{1}{2}\sum_{K=1}^{2P}\sum_{n_{K}=0}^{N}\sum_{n_{\overline{K}}=0}^{N}
\left[ n_{K}-n_{\overline{K}}\right] ^{2}P(n_{K},n_{\overline{K}})  \nonumber \\
&=&\frac{1}{2}\sum_{K=1}^{2P}\sum_{K^{\prime }=1}^{2P}\left\{
\sum_{n_{K}=0}^{N}\sum_{n_{K^{\prime }}=0}^{N}\left[ n_{K}-n_{K^{\prime }}
\right] ^{2}P(n_{K},n_{K^{\prime }})\right\} f_{K^{\prime },\overline{K}}, 
\nonumber
\end{eqnarray}
where $f_{K^{\prime },\overline{K}}$ is the probability that $K^{\prime }$ is the anticorrelated
strategy to $K$ (i.e. $\overline K$) and is hence given by $f_{K^{\prime },\overline{K}}=\delta
_{K^{\prime },2P+1-K}$. This manipulation is exact so far.

A switch is now made to the popularity-labels $\{Q\}$. After
relabelling, we obtain:
\begin{equation}
\sigma^{2}=\frac{1}{2}\sum_{Q=1}^{2P}\sum_{Q^{\prime }=1}^{2P}\left\{
\sum_{n_{Q}=0}^{N}\sum_{n_{Q^{\prime }}=0}^{N}\left[ n_{Q}-n_{Q^{\prime }}
\right] ^{2}P(n_{Q},n_{Q^{\prime }})\right\} f_{Q^{\prime },\overline{Q}},
\end{equation}
where $f_{Q^{\prime },\overline{Q}}$\ is the probability that the
strategy with label $Q^{\prime }$ is anticorrelated to $Q$. Consider any particular
strategy which was labelled previously by $K$ and is now labelled by $Q$. Unlike the
case of the flat disorder matrix, it is \emph{not} guaranteed that this strategy's
anticorrelated partner will lie in position ${Q^\prime}=2P+1-Q$. All that can be said is that the
strategy $R$ has changed label from $K\rightarrow Q(K)$ while the anticorrelated strategy has changed
label from $\overline{K}
\rightarrow \overline{Q}(\overline{K})$ and that in general $\overline{Q}
\neq 2P+1-Q$. As stated earlier, we have as a result of the relabelling that
$n_{Q=1}>n_{Q=2}>n_{Q=3}>n_{Q=4}>...$. It can be assumed
that as a zeroth-order approximation the values of $n_{Q=1}$, $n_{Q=2}$,
$n_{Q=3,\text{...}}$ etc. are still sharply peaked around their mean values
obtained for the flat-matrix case, i.e. it is assumed that the probability
distribution $P(n_{Q(K)},n_{Q^{\prime }(\overline{K})})$ will be sharply peaked
around the $n_{Q(K)}$\ and $n_{Q^{\prime }(\overline{K })}$ values given by the
mean values for a flat matrix. As before, we will label
these values ${\overline{n_{Q}}}$ and ${{\overline {n_{Q^{\prime }}}}}$ where the intrinsic
dependence of $Q$ on $K$ has been dropped. Hence $P(n_{Q},n_{Q^{\prime }})=
\delta_{n_{Q},
{{\overline{n_{Q}}}}}\delta_{n_{Q^{\prime }},{{\overline {n_{Q^{\prime }}}}}}$ with $
{\overline{n_{Q}}}$ and ${{\overline {n_{Q^{\prime }}}}}$ given by the bin-counting method.
Substituting in 
$P(n_{Q},n_{Q^{\prime }})=
\delta_{n_{Q},
{{\overline{n_{Q}}}}}\delta_{n_{Q^{\prime }},{{\overline {n_{Q^{\prime }}}}}}$ gives
\begin{equation}
\sigma^{2}=\frac{1}{2}\sum_{Q=1}^{2P}\sum_{Q^{\prime }=1}^{2P}\left[ 
{\overline{n_{Q}}}-{{\overline {n_{Q^{\prime }}}}}\right] ^{2}f_{Q^{\prime }
\overline{Q}},  \label{fq}
\end{equation}
where the function $f_{Q^{\prime },\overline{Q}}$, which is the
probability that the strategy with label $Q^{\prime }$ is anticorrelated
to strategy ${Q}$, still needs to be specified.

So what should the form of $f_{Q^{\prime },\overline{Q}}$ be? In the limit of negligible matrix
disorder at small $m$, as studied in the previous section, it will be a delta-function at 
$Q^{\prime}=2P+1-Q\equiv \overline{Q}$ since in this case the rankings in terms of popularity and
virtual-points will once more be identical. Hence we recover the results of the previous section.
However, as the relative importance of disorder in $\Psi$ increases, the
delta-function form will no longer be appropriate. One approximate solution to this problem
is to assume that the resulting disorder in the popularity ranking (as compared to the
original virtual-point ranking) is so strong, that the probability that
$Q^{\prime }$ is anticorrelated to ${Q}$ becomes
\emph{independent} of the label
$Q^{\prime }$ and is hence given by $1/(2P)$. Hence the anticorrelated strategy to $Q$ could lie
{\em anywhere} in the strategy list $\{ Q^\prime\}=1,\dots 2P$. In this sense, this is the  opposite
limit to the delta-function case corresponding to the flat disorder matrix at small $m$.
Hence instead of being a delta-function at
$Q^{\prime }=2P+1-Q$, it now follows that $f_{Q^{\prime },\overline{Q}}$ has a `flat'
form given by $1/(2P)$. In this limiting case we have 
\begin{equation}
\sigma^{{\rm flat}\ f}=\frac{N}{\sqrt{3}.2^{(m+1)/2}}(1-2^{-2(m+1)})^{\frac{1}{2}}. 
\label{nonflatom}
\end{equation} Comparing this with Equation \ref{flatom} it can be seen that
\begin{equation}
\sigma^{{\rm flat}\ f}=\frac{1}{\sqrt{2}}\sigma^{{\rm delta}\ f}\approx
0.7\sigma^{{\rm delta}\ f}
\end{equation} for the case of $S=2$. The superscripts `flat $f$' and `delta $f$' used in the
previous section, therefore can be seen to refer to the form of the probability function
$f_{Q^{\prime },\overline{Q}}$. We emphasize that Equation
\ref{flatom} can be derived from Equation \ref{fq} by letting $f_{Q^{\prime },\overline{Q}}$\
take the delta-function form $\delta _{Q^{\prime },2P+1-Q}$ peaked at
$Q^{\prime }=2P+1-Q\equiv{\overline Q}$.

\subsection{Non-flat quenched disorder matrix $\Psi$, large $m$}
For larger $m$, the standard deviation in the number of agents in a given bin is now
similar to the mean value. (By large $m$ it is meant that the number of strategies is greater than
$N.s$, i.e. $2.2^{m}>N.s$). Furthermore, there tend to be either 
$0$ or $1$ agents in each box $(Q,Q^{\prime })$. In this limit, there will tend to be $O(N)$ crowds,
with each crowd having $O(1)$ agent.  Hence the popularity ordering is highly degenerate since
$n_Q=0,1$ for all $Q$. Since the anticorrelated strategy to $Q$ could be anywhere, we have
$f_{Q^{\prime },\overline{Q}}=1/(2P)$. Using Equation \ref{fq}
then gives
\begin{eqnarray}
\sigma^{2} &=&\frac{1}{2}\sum_{Q=1}^{2P}\sum_{Q^{\prime }=1}^{2P}\left[ 
\overline{n_{Q}}-\overline{n_{Q^{\prime }}}\right] ^{2}f_{Q^{\prime },
\overline{Q}} \\ &=&\sum_{Q=1}^{N}\left\{ \left[
(\overline{n_{Q}}=1)-(\overline{n_{Q^{\prime }}}=1)\right] ^{2}\frac{N}{2P}+\left[
(\overline{n_{Q}}=1)-(
\overline{n_{Q^{\prime }}}=0)\right] ^{2}\frac{2P-N}{2P}\right\} ,  \nonumber
\end{eqnarray}
where the sum is now performed over the $N$ strategies with one agent subscribed.
Simplifying this expression gives
\begin{eqnarray}
\sigma^{{\rm flat}\ f,\ {\rm high}\ m} &=&\left(\sum_{Q=1}^{N}\left[
(\overline{n_{Q}}=1)-(\overline{n_{Q^{\prime }}}=0)
\right] ^{2}\frac{2P-N}{2P}\right) ^{\frac{1}{2}}  \label{highm} \\
&=&(N.\frac{2P-N}{2P})^{\frac{1}{2}}  \nonumber \\
&=&{\sqrt{N}}(1-\frac{N}{2^{m+1}})^{\frac{1}{2}},  \nonumber
\end{eqnarray}
where $P\equiv 2^{m}$ has been used. 
We note in passing that it is quite simple to obtain a slightly more sophisticated approximation for
$f_{Q^{\prime },\overline{Q}}$ which incorporates the agents' behavior in picking their most
successful strategy. For general $S$, this then yields
\begin{equation}
\sigma^{{\rm flat}\ f,\ {\rm high}\ m}={\sqrt{N}}(1-\frac{NS-1}{2^{m+1}})^{\frac{1}{2}}\ \ .
\end{equation}
[N.B. This latter expressions gives an even better fit to the numerical results shown later in Figure
7].

\subsection{Reduced vs. Full Strategy Space}
The general dynamics for a typical game are likely to be similar in character when played in both
the FSS and the RSS \cite{RSS}. Hence the RSS-based
Crowd-Anticrowd theory would generally be expected to provide a valid description for games played in
the FSS as well. In addition to straightforward numerical demonstration for the particular game of
interest, there is a theoretical justification which goes as follows. For a game played in the FSS
there are
$2^{2^{m}}/2.2^{m}$ distinct subsets of strategies. Each subset can be considered as a separate RSS.
Note that the strategies that belong to a given RSS are optimally spread out across the
corresponding FSS hypercube. Figure 2 shows the distribution of the $16$, $m=2$
strategies across the $4$ dimensional hypercube. The positions of the strategies
belonging to the RSS are such that no two strategies have a Hamming separation less
than $2^{m}/2$. The same can be said for any other choice of RSS. Due to the nature
of a RSS, and given similar strings of past outcomes over which to score strategies, each strategy
within the RSS attains a score in an uncorrelated or anticorrelated manner to any other strategy in
the subset. Any other RSS within the FSS will score its strategies in a similar way, although slightly
out of phase. For example for $m=3$, the first RSS to be considered could contain the strategy
$00001111$, and the second RSS considered could contain the strategy $01001111 $. It
is easy to see that on $7$ out of $8$ occasions these two strategies would score in
the same way. Hence it can be seen
that these two strategies from two separate RSS, will follow each other during a typical
run of an unbiased game. This argument extends across all strategies in all of the
$2^{2^{m}}/2.2^{m}$ distinct RSS's within the FSS. Hence a game using the FSS
behaves as if there are $2.2^{m}$ clusters of strategies and so is similar to a game
played in the RSS. These clusters form the Crowds and Anticrowds of the theory, and
this clustering allows the use of just one RSS in the theory.

\section{Crowd-Anticrowd theory applied to generalized Minority Games}
In this section we show that the present Crowd-Anticrowd theory, even within its RSS formulation,
provides a quantitative theory for both the Minority Game and several variants. 
There is currently no other analytic theory which can match the quantitative agreement
of the Crowd-Anticrowd theory over such a parameter range, and over such a range of game variants.
This justifies our belief that the Crowd-Anticrowd theory is an important milestone in such
multi-agent systems, and possibly even for Complex Systems in general. It is also pleasing from the
point of view of physics methodology, since the basic underlying philosophy of accounting correctly for
`inter-particle' correlations is already known to be successful in more conventional areas of
many-body physics. This success in turn raises the intriguing possibility that conventional many-body
physics might itself be open to re-interpretation in terms of an appropriate multi-particle `game': we
leave this for future work.

\subsection{Basic Minority Game}
Figure 7 shows the standard deviation $\sigma$ of the excess demand  $D[t]$, obtained for the
particular case of the Minority Game. The analytic
results were taken from Sections VA, VB and VC. The results are shown as a function of agent memory
size
$m$. The spread in numerical values from individual runs, for a given $m$, indicates the
extent to which the choice of $\Psi
$\ alters the dynamics of the MG. The upper line for each $S$ value at low $m$, is Equation
\ref{flatom}\ showing\
$\sigma^{{\rm delta}\ f}$. The lower line for each $S$ value at low $m$, is Equation
\ref{nonflatom}\ showing\ $\sigma^{{\rm flat}\ f}$. The line at high $m$ is
Equation \ref{highm}
\ showing $\sigma^{{\rm flat}\ f,\ {\rm high}\ m}$, and is independent of $S$. Comparing the analytic
curves with the numerical results, it can be seen that the analytic expressions
capture the essential physics (i.e. the strong correlations) driving the fluctuations in 
the system.

\subsection{Alloy Minority Game}
Here we consider the Crowd-Anticrowd theory applied to a mixed population containing
agents of different memory-sizes (or equivalently, agents with differing opinions as to the relative
importance of previous outcomes) \cite{alloy}. Since we are interested in the effects of crowding
within this mixed-ability population, we will focus on small $m$. Consider a population containing
some agents with memory
$m_{1}$, and some agents with memory $m_{2}$ where $m_1<m_2$. For a pure
population of agents with the same memory $m_1 $, there is information left
in the history time-series \cite{SavitLet}. In the small $m_1$ limit, however, this information is
hidden in bit-strings of length greater than $m_1$ and hence is not accessible to these
agents. However it would in principle be accessible to agents with a larger memory $m_2$.

In the mixed population or `alloy', there are two sub-populations comprising 
$N_{m_{1}}$ agents with memory $m_{1}$ and $S_1$ strategies per agent,
and $N_{m_{2}}=N-N_{m_{1}}$ agents with memory $m_{2}$ and $S_2$ strategies per agent. Let us focus on
the variance
$\sigma^2$ of demand
$D[t]$. If each population were pure, it would form Crowd-Anticrowd groups as discussed earlier in
this paper, and hence provide a variance given by the sum of the variances of these uncorrelated
groups. In the mixed population, we can assume as a first approximation that the individual
groups for different $m$ are also uncorrelated. In short, the agents looking at patterns of length
$m_1$, and the agents looking at patterns of length $m_2$, act in an uncorrelated way with respect to
each other. Hence the variance in the total $D[t]$ from both sub-populations, can be obtained by
adding separately the contributions to the variance from the $m_1$ agents and the
$m_2$ agents. Hence $\sigma ^{2}=\sigma
_{1}^{2}+\sigma _{2}^{2}$, where $\sigma _{1}$ ($\sigma _{2}$) is the
variance due to the $m_1$ ($m_2$) agents. Defining the concentration of $m_1$ agents as $x=N_{m_1}/N$,
gives $\sigma^2_{1}= C^2({m_1,S_1})x^2 N^2$ and
$\sigma^2_{2}= C^2({m_2,S_2})(1-x)^2 N^2$ where the expressions for $C^2({m_1,S_1})$ and
$C^2({m_2,S_2})$ follow from Equations \ref{final} and \ref{YofR}:
\begin{eqnarray}
C^2({m_i,S_i}) &=&\sum_{K=1}^{2^{m_i}}
\bigg(\bigg[1-\frac{(K-1)}{2^{m_i+1}}\bigg]^{S_i}-\bigg[1-
\frac{K}{2^{m_i+1}}\bigg]^{S_i} \\
&\ &\ \ -\ \bigg[1-\frac{(2^{m_i+1}-K)}{2^{m_i+1}}\bigg]^{S_i}+\bigg[1-
\frac{2^{m_i+1}+1-K}{2^{m_i+1}}\bigg]^{S_i}\bigg)^2   \nonumber
\end{eqnarray}%
Hence 
\begin{equation}
\sigma=N[C^{2}({m_1,S_1})x^{2}+C^{2}({m_2,S_2})(1-x)^{2}]^{1/2}\ . \label{alloy}
\end{equation}%
It can be seen that Equation \ref{alloy} will generally exhibit a {\em minimum} in $\sigma$ at finite
$x$, hence the mixed population uses the limited global resource more efficiently than a pure
population of either $(m_1,S_1)$ or $(m_2,S_2)$ agents. This analytic result has been confirmed in
numerical simulations of a mixed population \cite{alloy,dublin}. 

\subsection{Thermal Minority Game}
In the `Thermal
Minority Game' (TMG) \cite{SherThermLett}, agents choose between
their $S$ strategies using an exponential probability weighting. As pointed out
by Marsili \emph{et al} \cite{marsiliThermo}, such a probabilistic strategy
weighting has a long tradition in economics and encodes a particular
behavioral model. The numerical simulations of Cavagna \emph{et al}
demonstrated that at small $m$, where the MG $\sigma $ is
larger-than-random, the TMG $\sigma $ could be pushed below the random
coin-toss limit just by altering this relative probability weighting, or
equivalently the `temperature' $T$ \cite{SherThermLett}. This reduction in $%
\sigma $ for stochastic strategies seems fairly general: for example, in Ref. \cite{prl} we had
presented a modified MG in which agents with stochastic strategies also generate
a smaller-than-random $\sigma $. The common underlying phenomenon
in both cases is that stochastic strategy rules tend to
reduce (increase) the typical size of Crowds (Anticrowds), which in turn implies an {\em increase} in
the cancellation between the actions of the Crowds and Anticrowds. Hence 
$\sigma
$ gets reduced, and can even fall below the random coin-toss limit \cite{SherThermLett,prl}. 

We will now show that the Crowd-Anticrowd theory provides a quantitative explanation of the
main result of Ref. \cite{SherThermLett}, whereby a smaller-than-random $\sigma $ is generated with
increasing `temperature' \cite{generalTherm,CAA} at small $m$. We therefore focus on small $m$, and
assume a nearly flat strategy allocation matrix $\Psi$. At any moment in the game, strategies can be
ranked according to their virtual points, $K=1,2\dots 2^{m+1}$ where $K=1$ is the best strategy, $K=2$
is second best, etc. Consider any two strategies ranked $K$ and
$K^{\ast }$ within the list of $2^{m+1}$ strategies in the RSS. As mentioned earlier, 
in the small $m$ regime of interest the virtual-point
strategy ranking and popularity ranking for strategies are essentially the same. Consider $S=2$ as an
example.  Let
$p(K,K^{\ast }|K^{\ast }\geq K)$ be the probability that a given agent possesses $K$ and $K^{\ast}$,
where $K^{\ast }\geq K$ (i.e. $K$ is the best, or equal best, among his $S=2$ strategies). In
contrast, let $p(K,K^{\ast }|K^{\ast }\leq K)$ be the probability that a given agent possesses $K$ and
$K^{\ast }$, where $K^{\ast }\leq K$ (i.e. $K$ is the worst, or equal worst, among his $S=2$
strategies). Let $\theta$ be the probability that the agent uses the worst of his $S=2$ strategies,
while $1-\theta $ is the probability that he uses the best. The probability that the agent plays $K$
is then given by 
\begin{eqnarray}
p_{K} &=&\sum_{K^{\ast }=1}^{2^{m+1}}[\ \theta \ p(K,K^{\ast }|K^{\ast }\leq
K)+\ (1-\theta )\ p(K,K^{\ast }|K^{\ast }\geq K)]  \label{psubK} \\
&=&\ \theta \ p_{-}(K)+2^{-2(m+1)}\ \theta +\ (1-\theta )\ p_{+}(K)  \nonumber
\end{eqnarray}%
where $p_{+}(K)=\sum_{K^{\ast }}p(K,K^{\ast }|K^{\ast }\geq K)$ is the
probability that the agent has picked $K$ \emph{and} that $K$ is the agent's
best (or equal best) strategy; $p_{-}(K)=\sum_{K^{\ast }}p(K,K^{\ast
}|K^{\ast }<K)$ is the probability that the agent has picked $K$ \emph{and}
that $K$ is the agent's worst strategy. Using Equation \ref%
{YofR}\ it is straightforward to show that 
\begin{equation}
p_{+}(K)=\bigg(\bigg[1-\frac{(K-1)}{2^{m+1}}\bigg]^{2}-\bigg[1-\frac{K}{%
2^{m+1}}\bigg]^{2}\bigg)\ \ .  \label{pPlus}
\end{equation}%
Note that $p_{+}(K)+p_{-}(K)=p(K)$ where 
\begin{equation}
p(K)=2^{-m}(1-2^{-(m+2)})  \label{pK}
\end{equation}%
is the probability that the agent holds strategy $K$ after his $S=2$ picks,
with no condition on whether it is best or worst. An expression for $%
p_{-}(K) $ follows from Equations \ref{pPlus} and \ref{pK}. The basic MG
corresponds to the case $\theta =0$.

In the TMG, each agent is equipped at each timestep with his own (biased)
coin characterised by exponential probability weightings \cite{SherThermLett,thermnote}. An agent
then flips this coin at each timestep to decide which strategy to use \cite{SherThermLett}. To relate
the present analysis to the TMG in Ref. 
\cite{SherThermLett}, $0\leq \theta \leq 1/2$ is considered: $\theta =0$
corresponds to `temperature' $T=0$ while $\theta \rightarrow 1/2$
corresponds to $T\rightarrow \infty $ \cite{thermnote} with $\theta =1/2[1-\mathrm{tanh}%
(1/T)]$. Consider the mean number of agents playing strategy $K$, which is
now given by 
\begin{equation}
{\overline {n_{K}}}=Np_{K}=N\ (1-2\theta )\ p_{+}(K)+N\ \theta \ p(K)+2^{-2(m+1)}\ N\
\theta \ \ .  \label{nK}
\end{equation}%
Recall Equation \ref{final} 
\begin{equation}
\sigma_{\theta }=\bigg[\sum_{K=1}^{P}[{{\overline {n_{K}}}}-{{\overline {n_{\overline
K}}}}]^{2}\bigg]^{\frac{1}{2}} 
\end{equation}%
The
quantities 
${\overline {n_{K}}}$ and ${\overline {n_{\overline K}}}$ are now $\theta $-dependent (see Equation
\ref{nK}). Realizing that only the first term in Equation \ref{nK}\ is actually a
function of $K$ and hence substituting Equations \ref{pPlus}, \ref{pK} and %
\ref{nK} for $K$ and $\overline K$ into Equation \ref{final} yields 
\begin{equation}
\sigma _{\theta }=|1-2\theta |\ \sigma _{\theta =0}  \label{SigTheta}
\end{equation}%
where $\sigma _{\theta =0}$ is the standard deviation when $\theta =0$, i.e.
the standard deviation of $D[t]$ obtained for the basic MG. Equation \ref{SigTheta} explicitly shows
that the standard deviation $\sigma _{\theta }$ \emph{decreases} as $\theta $ increases
(recall $0\leq \theta \leq 1/2$): in other words, the standard deviation
decreases as agents use their worst strategy with increasing probability. An
increase in $\theta $ leads to a reduction in the size of the larger Crowds
using high-scoring strategies, as well as an increase in the size of the
smaller Anticrowds using lower-scoring strategies, hence resulting in a more
substantial cancellation effect between the Crowd and the Anticrowd. As $%
\theta $ increases, $\sigma _{\theta }$ will eventually drop \emph{below}
the random coin-toss result at $\theta =\theta _{c}$ where 
\begin{equation}
\theta _{c}=\frac{1}{2}-\frac{\sqrt{N}}{2}\frac{1}{\sigma _{\theta =0}}\ \ .
\end{equation}%
We now switch to the popularity labels $\{Q\}$ in order to examine $\sigma$ in the two limits of
`delta $f$', and `flat $f$'. The delta-function distribution is $\delta _{Q^{\prime
},2^{m+1}+1-Q}$ which is peaked at $Q^{\prime }=2^{m+1}+1-Q$, while the flat distribution is given by
$1/(2P)=[2^{-(m+1)}]$. Employing these approximate
distributions once again, the resulting expressions for
$\sigma _{\theta }$ are 
\begin{equation}
\sigma _{\theta}^{{\rm delta}\ f}=(1-2\theta )\frac{N}{{\sqrt 3}
2^{\frac{m}{2}}}\bigg(1-2^{-2(m+1)}\bigg)^{\frac{1}{2}}  \label{Thetad}
\end{equation}%
and 
\begin{equation}
\sigma _{\theta}^{{\rm flat}\ f}=(1-2\theta
)\frac{N}{\sqrt{3}\ 2^{(m+1)/2}}\bigg(1-2^{-2(m+1)}\bigg)^{\frac{1}{2}}
\label{thetaf}
\end{equation}%
respectively. For the TMG, $\theta =(1/2)[1-\mathrm{tanh}(1/T)]$ although
we again emphasize that the Crowd-Anticrowd theory is not limited to the case
of `thermal' strategy weightings.

Figure 8 shows a comparison between the theory of Equations %
\ref{Thetad}, \ref{thetaf} and numerical simulation for various runs, $m=2$
and $S=2$. The theory agrees well in the range $\theta =0\rightarrow 0.35$
and, most importantly, provides a quantitative explanation for the
transition in $\sigma $ from larger-than-random to smaller-than-random as $%
\theta $ (and hence $T$) is increased. The numerical data for different runs
has a significant natural spread. Most of these data points do lie
in the region in between the two analytic curves, which act as approximate upper and
lower-bounds. Above $\theta =0.35$, the numerical data tend to flatten off while the
present theory predicts a decrease in $\sigma $ as $\theta \rightarrow 0.5$.
This is because the present theory averages out the fluctuations in
strategy-use at each time-step (Equation \ref{nK} only considers the mean
number of agents using a strategy of given rank). Consider $\theta =0.5$. For a
particular configuration of strategies picked at the start of the game, and
at a particular moment in time, the number of agents using each strategy is
typically distributed \emph{around} the mean value ${\overline {n_{K}}}=N2^{-(m+1)}$ given
by Equation \ref{nK}\ for $\theta =0.5$. The resulting distribution
describing the strategy-use is therefore non-flat. It is these fluctuations
about the mean values ${\overline {n_{K}}}$ and ${\overline{n_{\overline K}}}$ which give rise to a
non-zero
$\sigma
$. The Crowd-Anticrowd theory can be extended to account for the effect of
these fluctuations in strategy-use for $\theta \rightarrow 0.5$ in the
following way: All $N$ agents are randomly assigned $S=2$ strategies. To
represent a turn in the game, each agent flips a (fair) coin to decide which
of the two strategies is the preferred one. Having generated a list of the
number of agents using each strategy, $\sigma $ is then found in the usual
way by cancelling off crowds and anticrowds. A time-averaged value for $%
\sigma $ is then obtained by averaging over $100$ independent coin-flip
outcomes for the given initial distribution of strategies among agents $%
\Psi $. This procedure provides a semi-analytic calculation for the value
of $\sigma $ at $\theta =0.5$. It is also possible to perform
a fully analytic calculation of the average $\sigma _{\theta }$ in the $%
\theta \rightarrow 0.5$ limit: the initial random assignment of strategies
can be modelled using a random-walk. This yields an average value of $
[\overline{n_{K}}-{\overline{n_{\overline{K}}}}]^{2}$ (i.e. time-averaged over the fluctuations)
given by $N2^{-m}(1-2^{-(m+1)})$. Summing over all pairs of
correlated-anticorrelated strategies, i.e. all Crowd-Anticrowd pairs, yields
a (configuration-average) standard deviation given by 
\begin{equation}
\sigma _{\theta \rightarrow 0.5}={\sqrt{N}}\big[1-2^{-2(m+1)}\big]^{%
\frac{1}{2}}\ \ .  \label{thetahalf}
\end{equation}%
The agreement for $\theta \rightarrow 0.5$ is good. Note that at finite $m$,
the value of $\sigma _{\theta \rightarrow 0.5}$ will always lie \emph{below} the random coin-toss
limit
$\sqrt{N}$, regardless of the value of
$N$. Hence the theoretical and numerical results agree, but neither is equal to the random
coin-toss limit as
$\theta
\rightarrow 0.5$ (i.e. $T\rightarrow
\infty $). Interestingly, we can therefore conclude that the high-temperature limit in this system
does not strictly correspond to the random coin-toss limit, regardless of the value of  $N$.

\subsection{Thermal Alloy Minority Game}
A variation of the Thermal Minority Game is now considered in which a
concentration $q$ of agents employ a probabilistic strategy selection at
each turn of the game. In particular, these agents play their worst strategy with
probability $%
\theta $, and hence play their best strategy with probability $(1-\theta )$.
These $qN$ agents will be called `TMG agents' because of the direct
connection with the Thermal Minority Game discussed above. The remaining $(1-q)N$ agents choose
their best strategy with probability unity (i.e. $\theta =0$ as in the basic MG) hence
they will be called `MG agents'. Numerical simulations show that as the
concentration $q$ of TMG agents increases, or the probability $\theta $
(i.e. $T$) increases, the standard deviation $\sigma $ decreases. We will now give the analytic
explanation of this in terms of Crowd-Anticrowds.

Recall from Equation \ref{psubK}, the probability that the agent plays the $K$'th
highest scoring strategy is given by 
\begin{eqnarray}
p_{K}^{\mathrm{TMG}} &=&\sum_{K^{\ast }=1}^{2^{m+1}}[\ \theta \ p(K,K^{\ast
}|K^{\ast }\leq K)+\ (1-\theta )\ p(K,K^{\ast }|K^{\ast }\geq K)]
\label{pkTMG} \\
&=&\ \theta \ p_{-}(K)+2^{-2(m+1)}\ \theta +\ (1-\theta )\ p_{+}(K)  \nonumber
\end{eqnarray}%
Setting $\theta =0$\ in Equation \ref{psubK}\ gives the probability that an
MG agent plays $K$ 
\begin{equation}
p_{K}^{\mathrm{MG}}=p_{+}(K)\ \ .
\end{equation}%
The mean number of agents $\overline {n_{K}}$ playing strategy $K$ in the
mixed-population game containing a concentration $q$ of TMG agents and $%
(1-q) $ of MG agents is now considered. This is given by 
\begin{eqnarray}
{\overline {n_{K}}} &=&\ q\ N\ p_{K}^{\mathrm{TMG}}+\ (1-q)\ N\ p_{K}^{\mathrm{MG}}
\label{nkall} \\
&=&\ N\ (1-2\ q\ \theta )\ p_{+}(K)+N\ q\ \theta \ p(K)+2^{-2(m+1)}\ N\ q\
\theta \ \ .  \nonumber
\end{eqnarray}%
Following the same analysis as in the previous section for the Thermal Minority Game, we obtain 
\begin{equation}
\sigma (q,\theta)=[1-2\ q\ \theta ]\ \{\sigma (q,\theta )\}_{q\theta =0}
\label{Sigqtheta}
\end{equation}%
where $\{\sigma (q,\theta )\}_{q\theta =0}$ is just the standard deviation
for the basic MG (i.e. $q=0$ and/or $\theta =0$). Equation \ref{Sigqtheta} predicts that the effect
on the standard deviation caused by a change in population composition and/or `temperature'
can be described by a simple prefactor $[1-2q\theta ]$. Provided that the
basic MG is in the crowded regime as discussed earlier, Equation \ref%
{Sigqtheta} holds for all $N$ and $m$ and hence any value of $\{\sigma
(q,\theta )\}_{q\theta =0}$. Hence a critical value $q_{c}$ can be
predicted for fixed $\theta $, or $\theta _{c}$ for fixed $q$, at which $%
\sigma (q,\theta )$ crosses from worse-than-random to better-than-random.
For a given value of $\theta $, it follows from Equation \ref{Sigqtheta}
that 
\begin{equation}
q_{c}(\theta )=\frac{1}{2\theta }-\frac{\sqrt{N}}{2\theta }\frac{1}{\{\sigma
(q,\theta )\}_{q\theta =0}}\ .  \label{qcrit}
\end{equation}%
A similar expression follows for $\theta _{c}(q)$. Given that $0\leq \theta
\leq 1/2$, Equation \ref{qcrit} implies that the run-averaged numerical
volatility should lie above the random coin-toss value if $q<q_{c}^{\ast }$
where 
\begin{equation}
q_{c}^{\ast }=1-\sqrt{N}\frac{1}{\{\sigma (q,\theta )\}_{q\theta
=0}}\ ,
\end{equation}%
regardless of `temperature' $T$. Since the $N$ and $m$ values
considered are such that the basic MG is in the worse-than-random regime, $%
\{\sigma (q,\theta )\}_{q\theta =0}\geq \sqrt{N}$ and therefore $0\leq
q_{c}^{\ast }\leq 1$ as required. Similarly $\sigma (q,\theta )$ will remain
above the random coin-toss value for all $q$ if $\theta <\theta _{c}^{\ast }$
where 
\begin{equation}
\theta _{c}^{\ast }=\frac{1}{2}-\frac{\sqrt{N}}{2}\frac{1}{\{\sigma
(q,\theta )\}_{q\theta =0}}\ .
\end{equation}

\section{B-A-R systems with an underlying network structure}
We now extend the above analysis to the case of a network connection between agents. 
In the first subsection, we consider the appropriate modification of the $\overline {n_K}$ values. In
the second sub-section, we follow the spirit of the alloy game mentioned above, whereby we consider the
case of very few connections within the population in order to show that in principle the standard
deviation can actually decrease as network connections are added.

\subsection{Modification of $\overline {n_K}$}
The presence of a network allows for a sharing of information across that network. Depending on the
rules of the game regarding information exchange, the connected agents may decide to adopt the
strategy prediction of agents to whom they are connected. The Crowd-Anticrowd
calculation can be generalized to incorporate such situations by forming new expressions for
$\overline {n_K}$. In particular, the bin-counting method to give
$\overline {n_K}$ must now be generalized to account for (i) any agent whose own strategies 
are lower ranking than $K$, but who is connected to another agent holding strategy
$K$, and (ii) any agent whose highest-scoring strategy is
$K$, but who has a connection to another agent with an even higher-scoring strategy.
The contribution (i) will increase $\overline {n_K}$ above the bin-counting value, however
contribution (ii) then reduces it. The competition between these two effects will determine what then
happens to the standard deviation in the presence of connections.

We will consider the following rule governing functionality of connections. Consider agent $i$
connected to agent $j$ in a game where each agent holds $S$ strategies (e.g. $S=2$). We also
suppose for the moment, that the ranking of strategies is unique (i.e. there are no strategies
which are tied in virtual-points).  Suppose that the highest-ranking strategy of agent $i$ is
$G$, but that the highest-ranking strategy of agent $j$ is
$K$ where $K<G$ and hence $K$ has higher virtual-point score than $G$. For the particular connection
rule we have in mind, we will let agent
$i$ therefore have access to the highest-scoring strategy of agent $j$. In other words, agent $i$ now
uses strategy
$K$ since it is the highest-scoring of the $2S$ strategies that the two agents hold between them. Of
course, agent $i$ may also be connected to other agents -- he will therefore use the highest-scoring
strategy among all the agents to whom he is connected. In the case that agent $j$ holds the
highest-scoring strategy of them all, then agent $i$ uses the strategy ranked $K$. The same is true
for all other agents. Hence we need to modify the calculation of ${\overline {n_K}}$ in order to
incorporate this effect. In particular, the number of agents using the $K$'th ranked strategy at a
particular timestep in the game-with-network will be
\begin{equation}
{\overline {n_K}}^{{\ \rm net}}={\overline {n_{K}}}+n_{\rightarrow
K}-n_{K\rightarrow}
\end{equation}
where $n_{\rightarrow K}$ is a sum over all agents who are connected to an
agent whose highest-ranking strategy is $K$, and who themselves have a highest-ranked strategy
$G$ which is worse than $K$ (i.e. $G$ is lower-ranked than $K$ and hence $G>K$). Hence these agents
use
$K$, whereas in the absence of the network they would have used their own strategy which is
lower-ranked than
$K$. By contrast,
$n_{K\rightarrow}$ is a sum over all agents whose highest-ranking strategy is
$K$, but who are connected to an agent whose highest-ranked strategy
$G$ is better than $K$ (i.e. $G$ is higher-ranked than $K$ and hence $G<K$). Hence these agents
use $G$, whereas in the absence of the network they would have used $K$. 
Notice that this is irrespective of the actual structure of the
network, for example the network could be random, small-world, scale-free, or regular.

In order to implement the renormalization of ${\overline {n_K}}$ in a particular example, we consider
the case of a random network in which agent $i$ has a probability of $p$ of forming a connection with
agent $j$. The term $n_{\rightarrow K}$
is given by summing over all agents whose own highest-scoring strategy $J$ is lower-ranked than $K$
(i.e.
$J>K$) {\em given} that they hold at least one connection to an agent whose highest-scoring strategy
is $K$ {\em but} they don't have any connections to any agents with a higher-ranked strategy $G$
(i.e.
 $G<K$). The resulting expression is
\begin{equation}
n_{\rightarrow K}=\bigg[\sum_{J>K} {\overline {n_J}}\bigg]\bigg[(1-p)^{\sum_{G<K}
{\overline {n_G}}}\bigg]\
\bigg[1-(1-p)^{{\overline {n_K}}}\bigg]
\end{equation}
where the third factor represents the probability that a given agent holds at least one connection to
an agent whose highest-scoring strategy is
$K$, the second factor accounts for the probability that a given agent is not connected to any
agent whose highest-scoring strategy $G$ is higher-ranked than $K$, and the first factor sums over
all agents whose highest-scoring strategy $J$ is lower-ranked than $K$. 
Continuing this analysis, we have
\begin{equation}
n_{K\rightarrow }={\overline {n_K}} \bigg[1-(1-p)^{\sum_{G<K} {\overline {n_G}}}\bigg]
\end{equation}
where the second factor accounts for the probability that a given agent has at least one connection
to an agent whose highest-scoring strategy $G$ is higher-ranked than $K$, and the first factor
is just the number of agents whose own highest-scoring strategy is $K$. 
Finally, we need a suitable expression ${\overline {n_K}}$, which is the number of agents using the
$K$'th highest-ranking strategy in the absence of the network.  As before, we will consider the case
of the small-$m$ limit with a flat strategy allocation matrix which, from Equation 23, gives 
\begin{equation}
{\overline {n_K}} = N\left( \left[
1-\frac{(K-1)}{2^{m+1}}\right] ^{S}-\left[ 1-
\frac{K}{2^{m+1}}\right] ^{S}\right) \ .
\end{equation}
These expressions can then be used to evaluate $\sigma$. Elsewhere we will discuss the
numerical results for a network B-A-R system \cite{charley}. In the limit of high connectivity $p$,
there is substantial crowding in the network B-A-R system and hence degeneracy of strategy scores.
Accounting for the correct frequency of degenerate/non-degenerate timesteps yields excellent
quantitative agreement with the numerical simulations \cite{charley}.

\subsection{Proof-in-principle that standard deviation can show a minimum at finite network
connectivity}
Here we will present a
proof-in-principle that the addition of connections can lead to a reduction in the standard
deviation $\sigma$ of demand $D[t]$. In fact there are two competing effects once more. First,
connections will tend to increase the size of crowds since more agents are now likely to play the
better-performing strategies. On the other hand, any particular agent who is connected to another,
will effectively have up to $2S$ strategies at his disposal, thereby increasing his chances of
uncovering a `good' strategy. It is this competition which interests us here.

We consider the interesting case in which the B-A-R system has very few connections. This might
correspond to a design problem in which initially there are no connections, and where the agents
themselves are already designed and built. Hence possible control of the system is limited
to introducing communication links between the agents. It is likely that such links will `cost' the
designer something, hence a cost-benefit question arises as to the extent to which the system can
be further modified or controlled by introducing a few connections into the system. In particular, how
many links should be added, and how should they be introduced? Here we will focus on a simpler version
of this problem: we assume that the links are expensive to introduce and cannot be formed in a
selective way. Hence we are assuming that there are very few links, and that the links that do exist
have been added in randomly to the system. The question we will address is then the subsequent effect
on the standard deviation of demand
$D[t]$ in the system. 

Suppose there is a given probability $p$ that any given agent $i$ is connected to agent $j$. The
population of agents therefore will contain, in general, a certain number $x_1(p)$ of agents who
are unconnected (i.e. cluster-size $n=1$), a number $x_2(p)$ of pairs of connected agents (i.e. 
cluster-size $n=2$), a number $x_3(p)$ of triples of connected agents (i.e. 
cluster-size $n=3$), etc. Hence we can effectively think of the population of $N$ agents as
comprising a `gas' containing $x_1(p)$ monomers, $x_2(p)$ dimers, $x_3(p)$ trimers, and
hence $x_n(p)$ $n$-mers where $n=1,2,3\dots$.
Hence we can write
\begin{equation}
N=\sum_{n=1}^N n x_n(p)
\end{equation}
For very low network connection probability, any particular
realization of the connections will result in most agents remaining unconnected while a few are
connected in pairs. Figure 9 shows a schematic diagram of this situation, whereby the population just
comprises monomers (i.e. isolated agents) and dimers (i.e. connected pairs of agents). This
implies that
\begin{equation}
N=x_1(p)+ 2 x_2(p) + \dots \approx x_1(p)+ 2 x_2(p)
\end{equation}
The mean number of dimers is simply given by the total number of possible pairs, i.e.
$\frac{1}{2}N(N-1)$, multiplied by the probability $p$ that a given pair is connected, hence
yielding $x_2(p)=\frac{1}{2} p N (N-1)$. Hence $x_1(p)=N-2.\frac{1}{2} p N (N-1)=N - p N (N-1)$. We
note that we are implicitly working below the percolation threshold: when $p\sim 1/N$ we have
that every agent has on average one connection, in which case the assumption of truncating at
dimers breaks down. 
We now turn to the calculation of the variance of $D[t]$. Recall that in the small $m$ limit with
flat matrices, the variance is given by
\begin{equation}
\sigma^2=\sum_{K=1}^{2^m} [\overline{n_{K}}-\overline{n_{\overline K}}]^2\ .
\end{equation}
For the case of no network (i.e. $p=0$), this becomes $\sigma_{p=0}^2 (m,S)=C^2(m,S) N^2$ where
we have added in the explicit dependence on $m$ and $S$, and have used the function defined earlier
on in Section VIB
\begin{eqnarray}
C^2({m,S_i}) &=&\sum_{K=1}^{2^{m}}
\bigg(\bigg[1-\frac{(K-1)}{2^{m+1}}\bigg]^{S_i}-\bigg[1-
\frac{K}{2^{m+1}}\bigg]^{S_i} \\
&\ &\ \ -\ \bigg[1-\frac{(2^{m+1}-K)}{2^{m+1}}\bigg]^{S_i}+\bigg[1-
\frac{2^{m+1}+1-K}{2^{m+1}}\bigg]^{S_i}\bigg)^2  \ \ . \nonumber
\end{eqnarray}%
Since the dimers are effectively `super-agents' with $2S$ strategies at their disposal, as opposed
to the isolated monomer agents who only have $S$, it is reasonable to suppose that the two
populations (i.e. monomers and dimers) are uncorrelated on average. This is similar in spirit to
the approximation used for the case of the alloy of mixed-memory agents in Section VI. Hence the
contribution to $D[t]$ by the monomer agents forms a stochastic process which is uncorrelated to the
contribution from the gas of dimer agents. Note that we are {\em not} assuming that the individual
monomers are not correlated with each other, or that the individual dimers are not correlated with
each other -- on the contrary, there will be significant crowding (i.e. correlations) within each
sub-population. The variance of
$D[t]$ for the combined population, can hence be written as follows
\begin{equation}
\sigma_p^2=\sum_{n=1}^{N} \sigma_{n;p}^2 = \sigma_{1;p}^2 + \sigma_{2;p}^2 + \dots
\end{equation}
The partial contributions are given as follows:
\begin{equation}
\sigma_{1;p}^2=[x_1(p)]^2 C^2(m,S)=[1-p(N-1)]^2 N^2 C^2(m,S)
\end{equation}
and
\begin{equation}
\sigma_{2;p}^2=2^2[x_2(p)]^2 C^2(m,2S)=[p(N-1)]^2 N^2 C^2(m,2S)
\end{equation}
where the factor $2^2$ reflects the fact that the dimer acts with step-size $2$ as opposed to $1$
for the monomers. Hence we obtain
\begin{equation}
\sigma_{p}^2=[1-p(N-1)]^2 \sigma_{p=0}^2(m,S) + [p(N-1)]^2 \sigma_{p=0}^2(m,2S)\ \ .
\end{equation}
This expression can then be minimized with respect to $p$, yielding a minimum in $\sigma_{p}^2$ at
\begin{equation}
p_{\rm min}=\bigg[(N-1)\bigg[1+\frac{\sigma_{p=0}^2(m,2S)}{\sigma_{p=0}^2(m,S)}\bigg]\bigg]^{-1}\ \ .
\end{equation}
Since $\sigma_{p=0}^2(m,2S)$ is larger than $\sigma_{p=0}^2(m,S)$ for small $m$, we have that
$p_{\rm min}<[N-1]^{-1}$. Hence we have proved that, to the extent to which a particular numerical
implementation reflects the connection rules assumed in the present analysis, adding in a small number
of connections can actually reduce the fluctuations in $D[t]$ and hence improve perfomance, in the
sense that the fluctuations in excess demand $D[t]$ are reduced. Interestingly, such a minimum in
$\sigma$ has already been reported based on numerical simulations for a somewhat similar game
\cite{zoltan}.  Our expression for $p_{\rm min}$ is likely to {\em underestimate} the actual
$p$ value at which a minimum occurs, since we have overestimated the coordination within a
given dimer, in addition to overestimating the number of dimers (and hence underestimating the effects
of trimers etc.). However the fact that the Crowd-Anticrowd analysis predicts that a minimum can in
principle exist, and Ref.
\cite{zoltan} had earlier found a minimum numerically in a similar game, is very encouraging.
 Clearly
there is a vast amount of further theoretical analysis that can be done within the present framework
-- however we leave this to future presentations.

\section{Conclusion}
We have given an in-depth presentation of the Crowd-Anticrowd theory in order to
understand the fluctuations in competitive multi-agent
systems, in particular those based on an underlying binary structure. Since the theory incorporates
details concerning the structure of the strategy space, and its possible coupling to history space, we
believe that the Crowd-Anticrowd theory will have applicability for more general multi-agent systems.
Hence we believe that the Crowd-Anticrowd concept might serve as a fundamental theoretical concept for
more general Complex Systems which mimic competitive multi-agent games. This would be a welcome
development, given the lack of general theoretical concepts in the field of Complex Systems as a
whole. It is also pleasing from the point of view of physics methodology, since the basic underlying
philosophy of accounting correctly for `inter-particle' correlations is already known to be successful
in more conventional areas of many-body physics. This success in turn raises the intriguing
possibility that conventional many-body physics might be open to re-interpretation in terms of an
appropriate multi-particle `game': we leave this for future work.

Some properties of multi-agent games cannot be described using
time- and configuration-averaged theories. In particular, an observation of a real-world Complex
System which is thought to resemble a multi-agent game, may correspond to a
\emph{single} run which evolves from a specific initial configuration of agents'
strategies. This implies a particular $\Psi$, and hence the time-averagings within the
Crowd-Anticrowd theory must be carried out for that particular choice of $\Psi$. However this problem
can still be cast in terms of the Crowd-Anticrowd approach, since the averagings are then just carried
out over some sub-set of paths in history space, which is conditional on the path along which the
Complex System is already heading. We also emphasize that a single $\Psi$ `macrostate' corresponds to
many possible `microstates', where each microstate corresponds to one particular partition
of strategy allocation among the agents. Hence the Crowd-Anticrowd theory retained at the level
of a given specified
$\Psi$, is equally valid for the entire {\em set} of games which share this same `macrostate'. We
refer to David Smith's presentation at this Workshop for a detailed discussion of $\Psi$-specific
dynamics. See also Refs.
\cite{THMG1,THMG2} for the simpler case of the Minority Game.

We have been discussing a Complex System based on multi-agent dynamics, in which both
deterministic and stochastic processes co-exist, and are indeed intertwined. Depending on the
particular rules of the game, the stochastic element may be associated with any of five areas: (i)
disorder associated with the strategy allocation and hence with the heterogeneity in the population,
(ii) disorder in the underlying network. Both (i) and (ii) might typically be fixed from the outset
(i.e., quenched disorder) hence it is interesting to see the interplay of (i) and (ii) in terms
of the overall performance of the system. The extent to which these two `hard-wired' disorders might
then compensate each other, as for example in the Parrondo effect or stochastic resonance, is an
interesting question. Such a compensation effect might be engineered, for example, by altering
the rules-of-the-game concerning inter-agent communication on the existing network. Three further
possible sources of stochasticity are (iii) tie-breaks in the scores of strategies, (iv) a stochastic
rule in order for each agent to pick which strategy to use from the available $S$ strategies, as in the
Thermal Minority Game, (v) stochasticity in the global resource level $L[t]$ due to changing external
conditions. To a greater or lesser extent, these five stochastic elements will tend to break up any
deterministic cycles arising in the game. In the case of (iii), the stochasticity due to tie-breaks,
we note that this stochasticity may actually have a physical significance. Suppose a real agent is at
a deadlock in terms of which strategy to use at a particular timestep: some
additional microscopic factor may then become important in determining the particular strategy that
the agent follows (e.g. power level in the case of mechanical agents) hence breaking the tie. In this
way,  the coin-toss mimics the effect of this additional microscopic characteristic of the
individual agents, which had been left out of the original game. 

The future therefore looks very interesting for these Binary Agent Resource models. By moving
beyond the confines of the El Farol Problem and Minority Games -- for example by introducing
underlying network structures and generalized game rules -- the possibility exists for pursuing a
general yet cohesive path through the Complex Systems landscape. On the control and design aspect, the
fascinating question now arises concerning the best way to insert the Crowd-Anticrowd theory within
the powerful COIN (Collective Intelligence) framework developed by Wolpert and Tumer for
Collectives\cite{davidnasa}. This question will be addressed in future works.

\vskip0.2in
\begin{acknowledgments}
We are grateful to Michael Hart and Paul Jefferies for their earlier input to the material in this
draft. One of us (PMH) acknowledges the support from the Research Grants Council  of the Hong Kong SAR
Government under Grant No. CUHK4241/01P.

\end{acknowledgments}

\newpage

{\bf FIGURES}

\vskip0.5in
\noindent {\bf FIGURE 1}  
Schematic representation of B-A-R (Binary Agent Resource) system. At
timestep
$t$, each agent decides between action $-1$ and action $+1$ based on the
predictions of the $S$ strategies that he possesses. A total of $n_{-1}[t]$ agents choose
 $-1$, and $n_{+1}[t]$ choose $+1$. In the simplified case 
that each agent's confidence threshold for entry into the game is very small, then
$n_{-1}[t]+n_{+1}[t]=N$ (i.e. all agents play at every timestep). Agents may be subject to
some underlying network structure which may be static or evolving, and ordered or disordered. The
strategy allocation, and hence heterogeneity in the population, provides a further source of disorder
which may be static (i.e. quenched) or evolving. The algorithm for deciding the
global outcome, and hence which action was winning/losing, lies in the hands of a
ficticious `Game-master'. This Game-master aggregates the agents' actions and then announces the
global outcome. All strategies are then rewarded/penalized according to whether
they had predicted the winning/losing action.

\vskip0.5in
\noindent {\bf FIGURE 2}
Strategy Space for $m=2$, together with some example strategies
(left). The strategy space shown is known as the Full Strategy Space FSS, and
contains all possible permutations of the actions $-1$ and $+1$ for each
history. There are $2^{2^{m}}$ strategies in the FSS. The $2^{m}$ dimensional
hypercube (right) shows all 
$2^{2^{m}}$ strategies from the FSS at its vertices. The shaded strategies form a Reduced Strategy
Space RSS. There are $2.{2^{m}}=2^{m+1}$ strategies in the RSS. The red shaded line connects two
strategies with a Hamming distance separation of 4.

\vskip0.5in
\noindent {\bf FIGURE 3}
History Space. Examples of the de Bruijn graph for $m=1,2,$ and $3$. Red transitions between
states correspond to the most recent global outcome $0$. Blue transitions between
states correspond to the most recent global outcome $1$.

\vskip0.5in
\noindent {\bf FIGURE 4}
Example distribution for the tensor $\Omega$ describing the 
strategy allocation for
$N=101$ agents in the case of $m=2$ and $S=2$, 
for the reduced strategy space RSS.

\vskip0.5in
\noindent {\bf FIGURE 5}
Schematic diagram of a fairly typical variation in strategy scores, as a function of time,
for a competitive game. This behavior is particularly relevant for the low $m$ regime where there are
many more agents than strategies, and hence strategy rankings change in time due to being
overplayed. The strategies at any given timestep can be ranked in terms of virtual-point
ranking
$K$, with
$K=1$ as the highest-scoring and
$\overline {K=1}$ as the lowest-scoring. The actual identity of the strategy in rank $K$ changes 
as time progresses, as can be seen. Ignoring accidental ties in score, there is a well
defined ranking of strategies at each timestep in terms of their $K$ values.

\vskip0.5in
\noindent {\bf FIGURE 6}
Schematic representation of the strategy allocation matrix
$\Psi$ with $m=2$ and $s=2$, in the RSS. The strategies are ranked according to strategy score, and
are labelled by the rank $K$. In the limit that $\Psi$ is
essentially flat, then the number
of agents playing the $K$'th highest-scoring strategy, is just proportional to the number of shaded
bins at that $K$.

\vskip0.5in
\noindent {\bf FIGURE 7}
Crowd-Anticrowd theory vs. numerical simulation results for Minority Game 
as a function of memory size $m$, for $N=101$ agents, at $S=2$, $4$ and $8$.
At each $S$ value, analytic forms of standard deviation in excess demand $D[t]$, are shown
corresponding to 
$\sigma^{{\rm delta}\ f}$ (upper solid line),
$\sigma^{{\rm flat}\ f}$ (lower dashed line) and $\sigma^{{\rm flat}\ f,\ {\rm high}\ m}$
(monotonically-increasing solid line which is independent of $S$). The numerical
values were obtained from different simulation 
runs (triangles, crosses and circles). 

\vskip0.5in
\noindent {\bf FIGURE 8}
Crowd-Anticrowd theory vs. numerical simulation results for Thermal Minority Game 
as a function of stochastic probability $\theta$, or `temperature' $T$. The analytic results (lines)
correspond to the $\sigma _{\theta}^{{\rm delta}\ f}$ (solid upper line) and 
$\sigma _{\theta}^{{\rm
flat}\ f}$ (solid lower line) limiting-case approximations.

\vskip0.5in
\noindent {\bf FIGURE 9}
Schematic representation of B-A-R (Binary Agent Resource) network system in limit of few
connections.


\begin{thebibliography}{99}
\bibitem{thermnote} The Thermal Minority Game discussed in Ref. \cite {SherThermLett} depends on a
parameter
$T$ (or equivalently $1/\beta $) called a `temperature'. We could similarly define $T$ by setting the
probability of playing the worst strategy $\theta =e^{-\beta }/(e^{\beta }+e^{-\beta })$. Hence
$T=2[{\rm ln}(\theta ^{-1}-1)]^{-1}$. $T=0$ corresponds to $\theta =0$ while
$T\rightarrow \infty $ corresponds to $\theta \rightarrow 1/2$, hence we will only consider $0\leq
\theta \leq 1/2$.
\end{thebibliography}
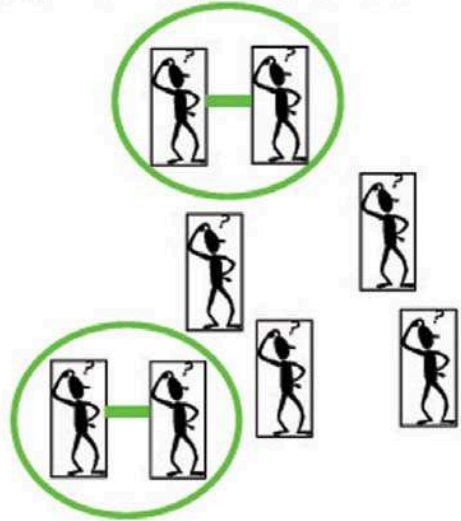
\end{document}